\documentclass[letterpaper,11pt]{article}
\def\anon{0}

\usepackage[utf8]{inputenc}
\usepackage[table,xcdraw]{xcolor}

\usepackage{tikz}
\usetikzlibrary{arrows, automata, arrows.meta, calc}
\usepackage{amsmath, amsthm, amssymb, thm-restate}
\makeatletter
\@ifundefined{newcounteralias}{}{%
  \renewcommand\thmt@autorefsetup{%
    \@xa\def\csname\thmt@envname autorefname\@xa\endcsname\@xa{\thmt@thmname}%
  }%
}
\makeatother
\usepackage{algorithmicx}
\usepackage[ruled,vlined,linesnumbered]{algorithm2e}
\usepackage{setspace}
\usepackage{mathtools}
\usepackage[numbers]{natbib}
\usepackage{comment} 
\usepackage[most]{tcolorbox} 
\usepackage{xfrac}
\usepackage{hyperref}
\usepackage{multirow}
\usepackage{caption}
\usepackage{bm}
\usepackage{float}
\usepackage{newfloat}
\usepackage{enumitem}
\usepackage{dblfloatfix} 
\usepackage{wrapfig}
\usepackage{tikz}
\usepackage{csquotes}
\usetikzlibrary{trees, shapes, decorations.pathreplacing, angles, quotes}
\usetikzlibrary{shapes.geometric, positioning, arrows.meta}

\usepackage{fancyhdr}

\SetCommentSty{mycommfont}
\let\oldnl\nl
\newcommand{\nonl}{\renewcommand{\nl}{\let\nl\oldnl}}

\usepackage[margin=1in]{geometry}
\usetikzlibrary{fit}
\allowdisplaybreaks

\definecolor{mygreen}{RGB}{10,120,240}
\definecolor{myred}{RGB}{10,120,240}

\hypersetup{
     colorlinks=true,
     citecolor= mygreen,
     linkcolor= myred,
     urlcolor=myred
}

\renewcommand{\epsilon}{\varepsilon}

\DeclareMathOperator{\E}{\ensuremath{\mathbb{E}}}

\newcommand{\hL}{\widehat{L}}

\newcommand{\hM}{\widehat{M}}
\newcommand{\hA}{\widehat{A}}
\newcommand{\hZ}{\widehat{Z}}
\newcommand{\hmu}{\widehat{\mu}}
\newcommand{\bb}{\bm{b}}
\newcommand{\htau}{\widehat{\tau}}
\newcommand{\Fdual}{F_{\mathrm{dual}}}
\newcommand{\Sbase}{S_{\mathrm{base}}}
\newcommand{\Ssplit}{S_{\mathrm{split}}}
\newcommand{\Sroot}{S_{\mathrm{root}}}
\newcommand{\Froot}{F_{\mathrm{root}}}
\newcommand{\Fsplit}{F_{\mathrm{split}}}
\DeclarePairedDelimiter{\hull}{\langle}{\rangle}
\newcommand{\RC}{\mathcal{R}(C)}
\newcommand{\RT}{\mc{R}(T)}
\newcommand{\imax}{i_{\max{}}}
\newcommand{\U}{\mc{U}}
\newcommand{\X}{\mc{X}}

\newcommand{\hiddencomment}[1]{}

\newcommand{\mc}[1]{\ensuremath{\mathcal{#1}}}

\newcommand{\MST}{\mathrm{MST}}
\newcommand{\TSP}{\mathrm{TSP}}

\DeclarePairedDelimiter\card{\lvert}{\rvert}

\newcommand{\floor}[1]{\left\lfloor #1 \right\rfloor}
\newcommand{\ceil}[1]{\left\lceil #1 \right\rceil}

\newcommand{\Ot}{\ensuremath{\widetilde{O}}}

\newcommand{\Omegat}{\ensuremath{\widetilde{\Omega}}}

\usepackage[noabbrev,nameinlink]{cleveref}
\crefname{lemma}{Lemma}{Lemmas}
\crefname{theorem}{Theorem}{Theorems}
\crefname{property}{Property}{Properties}
\crefname{claim}{Claim}{Claims}
\crefname{definition}{Definition}{Definitions}
\crefname{observation}{Observation}{Observations}
\crefname{proposition}{Proposition}{Propositions}
\crefname{assumption}{Assumption}{Assumptions}
\crefname{line}{Line}{Lines}
\crefname{figure}{Figure}{Figures}
\creflabelformat{property}{(#1)#2#3}
\crefname{equation}{}{}
\crefname{section}{Section}{Sections}
\crefname{appendix}{Appendix}{Appendices}
\crefname{algCounter}{Algorithm}{Algorithms}
\Crefname{algCounter}{Algorithm}{Algorithms}

\newtheorem{lemma}{Lemma}[section]
\newtheorem{theorem}{Theorem}

\newtheorem{corollary}[lemma]{Corollary}

\newtheorem{definition}[lemma]{Definition}
\newtheorem{claim}[lemma]{Claim}

\counterwithin{figure}{section}

\definecolor{mylightgray}{RGB}{240,240,240}

\algnewcommand{\IIf}[2]{\textbf{if} #1 \textbf{then} #2}
\algnewcommand{\EndIIf}{\unskip\ \algorithmicend\ \algorithmicif}
\algnewcommand{\IElse}[1]{\textbf{else} #1}

\newenvironment{graytbox}{
\par\addvspace{0.1cm}
\begin{tcolorbox}[width=\textwidth,
                  boxsep=5pt,
                  left=1pt,
                  right=1pt,
                  top=2pt,
                  bottom=2pt,
                  boxrule=1pt,
                  arc=2pt,
                  colback=mylightgray,
                  colframe=black,
                  ]
}{
\end{tcolorbox}
}

\newenvironment{whitetbox}{
\par\addvspace{0.1cm}
\begin{tcolorbox}[width=\textwidth,
                  boxsep=5pt,
                  left=1pt,
                  right=1pt,
                  top=2pt,
                  bottom=2pt,
                  boxrule=1pt,
                  arc=0pt,
                  colframe=black,
                  colback=white
                  ]
}{
\end{tcolorbox}
}

\newcounter{algCounter}

\let\oldlemma\lemma
\renewcommand{\lemma}{%
  \renewcommand{\emph}[1]{\textbf{##1}}
  \oldlemma}

\let\olddefinition\definition
\renewcommand{\definition}{%
  \renewcommand{\emph}[1]{\textbf{##1}}
  \olddefinition}

\usepackage{amsthm} 
\usepackage{etoolbox} 
\usepackage{comment} 

\ifnum\anon=0
\newcommand{\sbnote}[1]{{\color{blue} [Soheil: #1]}}
\newcommand{\mnote}[1]{{\color{red} [Madhu: #1]}}
\newcommand{\aanote}[1]{{\color{red} [Amir: #1]}}
\newcommand{\agnote}[1]{{\color{blue} [Alma: #1]}}
\else  
\newcommand{\aanote}[1]{}
\newcommand{\mnote}[1]{}
\newcommand{\agnote}[1]{}
\newcommand{\sbnote}[1]{}
\fi 

\makeatletter
\renewcommand{\paragraph}{%
  \@startsection{paragraph}{4}%
  {\z@}{10pt}{-1em}%
  {\normalfont\normalsize\bfseries}%
}
\makeatother

\makeatletter
\patchcmd{\@algocf@start}
  {-1.5em}
  {0pt}
  {}{}
\makeatother

\title{Metric TSP in Sublinear Time: Breaking Two Approximation}

\if\anon1{}\else{
\author{
Amir Azarmehr\thanks{Northeastern University, Boston, Massachusetts, USA.  Supported in part by NSF CAREER Award CCF 2442812, a Sloan Fellowship, and a Google Research Award. Emails: \texttt{\{azarmehr.a, s.behnezhad\}@northeastern.edu}.} \and 
Soheil Behnezhad\footnotemark[1]
}
}\fi 

\date{}

\begin{document}

\maketitle

\thispagestyle{empty}
\begin{abstract}

The seminal result of Czumaj and Sohler [STOC'04] showed that a 2-approximation of the value of metric TSP can be obtained with sublinear queries. Whether sublinear algorithms can break the 2-approximation barrier has remained a major open problem. See \href{https://sublinear.info/index.php?title=Open_Problems:71}{problem 71} on sublinear.info.

\smallskip\smallskip

This paper resolves this open problem. We show there are fixed $\varepsilon, \delta > 0$ such that a $(2-\delta)$-approximation of metric TSP value can be obtained with $O(n^{2-\varepsilon})$ queries. Here $n$ is the number of vertices, and the input size is $\Theta(n^2)$.

\end{abstract}

{
\clearpage
\hypersetup{hidelinks}
\vspace{1cm}
\renewcommand{\baselinestretch}{0.1}
\setcounter{tocdepth}{2}
\tableofcontents{}
\thispagestyle{empty}
\clearpage
}

\clearpage

\setcounter{page}{1}

\section{Introduction}
\label{sec:introduction}

We study the {\em metric traveling salesperson problem} (TSP) in the sublinear query model. 
Given an $n$-point metric $(V,d)$, let $\TSP$ denote the length of a shortest Hamiltonian cycle. Our goal is to estimate the value of $\TSP$ using a sublinear number of {\em queries}: a query
specifies two vertices $u,v\in V$ and returns $d(u,v)$.
Since the matrix has $\Theta(n^2)$ entries, sublinear query complexity
in this setting means making $o(n^2)$ distance queries.

The seminal work of Czumaj and Sohler~\cite{CzumajS04}
established that the weight of a metric minimum spanning tree (MST)
can be estimated within a factor of $1+\epsilon$ using
$\Ot(n/\epsilon^{O(1)})$ distance queries, for every $\epsilon>0$.
This immediately gives an approximation approaching a factor of $2$
for metric TSP given the well-known fact that
\[
    \MST \leq \TSP \leq 2\MST.
\]
The lower bound follows by deleting an edge of an optimal tour.
For the upper bound, double every edge of an MST, take an Euler tour,
and shortcut repeated vertices using the triangle inequality.

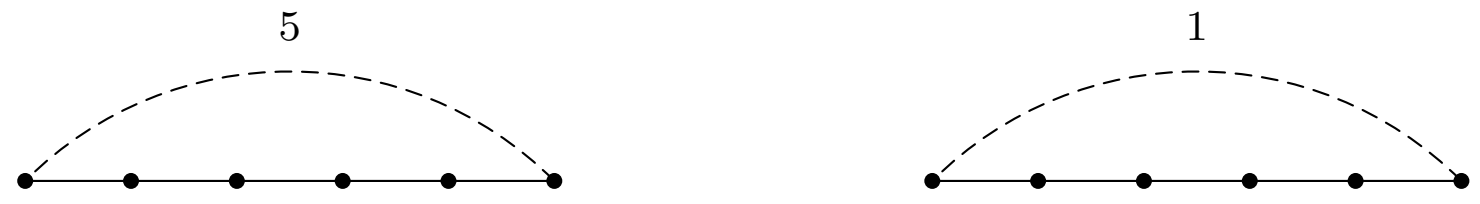
\begin{figure}[H]
    \centering
    \begin{tikzpicture}[
    line cap=round,
    every node/.style={font=\scriptsize}
]
    \begin{scope}
        \foreach \i in {0,...,4} {
            \draw ({0.7*\i},0) -- ({0.7*(\i+1)},0);
        }
        \foreach \i in {0,...,5} {
            \fill ({0.7*\i},0) circle (1.5pt);
        }
        \draw[densely dashed]
            (3.5,0) to[out=135,in=45]
            node[midway,above=2pt] {$5$} (0,0);

        \node[align=center] at (1.75,-0.65)
            {(a) Path metric\\[2pt]
             };
    \end{scope}

    \begin{scope}[xshift=6cm]
        \foreach \i in {0,...,4} {
            \draw ({0.7*\i},0) -- ({0.7*(\i+1)},0);
        }
        \foreach \i in {0,...,5} {
            \fill ({0.7*\i},0) circle (1.5pt);
        }
        \draw[densely dashed]
            (3.5,0) to[out=135,in=45]
            node[midway,above=2pt] {$1$} (0,0);

        \node[align=center] at (1.75,-0.65)
            {(b) Cycle metric\\[2pt]
             };
    \end{scope}
\end{tikzpicture}
    \caption{
        Solid edges form an MST and have unit weight;
        dashed edges indicate the metric distance between the endpoints of the path.
    }
    \label{fig:examples-of-K}
\end{figure}

As it turns out, breaking the factor-$2$ barrier requires substantially more information
about the structure of the metric than its MST weight.
The same exact MST can occur in metrics whose TSP values differ
by a factor of $2$.\footnote{\Cref{fig:examples-of-K} gives one such example.}
On top of this, the algorithm of \cite{CzumajS04} crucially only estimates the value of the MST and does not return its edges, as finding the edges of any constant approximate MST requires $\Omega(n^2)$ queries \cite{Indyk99}. This substantially complicates obtaining a better than two approximation of TSP. Despite significant interest \cite{ChenKK20,Beh21,BhattacharyaKS23,ChenKT23TSP,BehnezhadRRS24}, it remains a central
open problem after over two decades since the work of Czumaj and Sohler. See \href{https://sublinear.info/index.php?title=Open_Problems:71}{Open Problem~71} on sublinear.info for some more context about this problem.

\paragraph{Our result.}
We resolve this problem by giving an algorithm that estimates the value
of metric TSP within a factor strictly smaller than $2$ using a truly
subquadratic number of distance queries.

\begin{graytbox}
\begin{theorem}[Main result]
\label{thm:main}
There exist absolute constants $\epsilon,\delta>0$ and a randomized
algorithm that, given query access to the distance matrix of any
$n$-point metric, makes $O(n^{2-\epsilon})$ distance queries and, with probability
at least $2/3$, returns an estimate $\widehat{\TSP}$ satisfying
\[
    \TSP \leq \widehat{\TSP}
    \leq (2-\delta)\TSP.
\]
Additionally, the algorithm runs in $O(n^{2 - \gamma})$ time for an absolute constant $\gamma > 0$.
\end{theorem}
\end{graytbox}

The constants $\epsilon$ and $\delta$ are independent of $n$ and of
the input metric. The algorithm uses only distance queries, with no
MST supplied and no restriction to a special class of metrics.

\paragraph{Prior partial progress.}
Although the problem had remained open in general metrics prior to our work, significant progress was made on several special metrics of interest.

For \emph{$(1,2)$-metrics}, where every distance is either $1$ or $2$, an observation of Gupta and Onak~\cite{Onak16} 
together with the sublinear time matching estimator of Behnezhad~\cite{Beh21},
gives a $(1.75+\epsilon)$-approximation using $\Ot(n)$ distance queries.
Chen, Kannan, and Khanna~\cite{ChenKK20} obtained a
$(1.625+\epsilon)$-approximation using $\Ot(n^{3/2})$ distance queries
by estimating pairs of edge-disjoint matchings.
Behnezhad, Roghani, Rubinstein, and Saberi~\cite{BehnezhadRRS24}
developed a sublinear estimator for the maximum number of edges in
a collection of vertex-disjoint paths, obtaining a $(1.5+\epsilon)$-approximation for $(1,2)$-TSP using
$\Ot(n)$ distance queries.

For \emph{graphic metrics}---where distances are shortest-path distances over a connected, unweighted, undirected graph---
Chen, Kannan, and Khanna~\cite{ChenKK20} gave the first
$(2-\epsilon_0)$-approximation using $\Ot(n)$ distance queries, for an
absolute constant $\epsilon_0>0$, and a $1.93$-approximation
using $\Ot(n^{3/2})$ distance queries.
Combining the latter algorithm with the matching estimator of
Behnezhad~\cite{Beh21} reduces the number of distance queries to $\Ot(n)$.
The path-cover methods of \cite{BehnezhadRRS24}
yield a $1.84$-approximation using $\Ot(n)$ distance queries
and a $1.67$-approximation using
$n^{2-\Omega(1)}$ distance queries.

\paragraph{Unit-weight and supplied MSTs.}
The closest paper to this paper is that of Chen, Khanna, and Tan~\cite{ChenKT23TSP} who made progress toward
general metrics by showing that a $(2-\delta)$-approximation,
for some fixed constant $\delta>0$, could be obtained using
$\Ot(n^{3/2})$ queries under one of two different assumptions.
Working with metrics whose nonzero distances are positive integers,
their first result assumes that some MST consists entirely of edges
of weight~$1$; the tree itself need not be given.
This promise includes graphic metrics and
allows distances that are not determined by the graph of
weight-$1$ edges.
Their second result allows arbitrary MST edge weights, provided an
MST is given for free.
Their approach uses the \emph{cover advantage} of an MST to measure
potential savings over the doubled tree.
The restrictions leave unresolved how to obtain a strict improvement
over $2$ when the metric is unrestricted and the MST is unknown.
\Cref{thm:main} resolves this.

\paragraph{Lower bounds.} It is worth noting that Chen, Kannan, and Khanna~\cite{ChenKK20} showed
that, for some absolute constant $\delta>0$, a
$(1+\delta)$-approximation requires $\Omega(n^2)$ distance queries
even for $(1,2)$-metrics and for graphic metrics.
Thus, unlike MST, an arbitrary good approximation of metric TSP cannot be obtained with sublinear queries.

\section{Technical Overview}\label{sec:overview}

Given a metric $(V, d)$, we use $n = \card{V}$ to denote the number of vertices, $T$ to denote the minimum spanning tree, $M = w(T)$ to denote its weight, and $\TSP$ to denote the weight of the shortest Hamiltonian cycle.
We use $\Ot(\cdot)$ to suppress logarithmic factors in $n$.

\subsection{TSP through the Cut LP}

In this section we explore the main tool for our approximation, which is the Cut LP,
as well as a closely related problem that we call \emph{MST shortening}, which is connected to the Cut LP through its dual.
The next section explains how we can leverage these tools in a sublinear setting.

\subsubsection{The Cut LP}
At the heart of our algorithm is the cut-relaxation LP for MST,
which assigns fractional values to edges such that any cut is assigned a value of at least $1$:
\begin{equation}
K = \min\left\{\sum_e w_e x_e \ \middle\lvert\  x(\delta(A)) \geq 1\quad (\emptyset \neq A \subsetneq V), \quad x\geq 0\right\}. \tag{$\mathcal{P}$} \label{eq:primal}
\end{equation}
Here, the edge set consists of all vertex pairs, the weight of an edge $e = (u, v)$ is given by the distance between its endpoints $w_e = d(u, v)$, and $\delta(A)$ is used to denote the set of outgoing edges of $A$, that is, the cut associated with the vertex set $A$. Intuitively, the constraints of the LP ensure that the chosen subgraph has the minimum weight among all subgraphs that {\em fractionally} span the entire vertex set. It has to be noted that the Cut LP is different from the typical LP relaxation for MST which also has the additional constraint that $\sum_e x_e = n-1$ and $\sum_{e\subseteq A}x_e\leq |A|-1$ for every $A\subseteq V$. In fact, under metric weights, the Cut LP is equivalent to the well-known subtour-elimination LP up to rescaling \cite{DantzigFJ54,GoemansB93},
which has served as a key tool in TSP approximation (see \cite{KarlinKG21,KarlinKG22,GharanSS11,ChekuriQ17} for some examples, and \cite{traub2024approximation} for a comprehensive treatment).

To obtain our result, we reduce the better-than-2 approximation of TSP to the better-than-2 approximation of $K$. Two key inequalities allow us to do so. First,
\begin{equation}
\frac{M}{2} \leq K \leq M. \label{eq:K-bounds}
\end{equation}
Second,
\begin{equation}
\max(M, 2K) \leq \TSP \leq M + K. \label{eq:TSP-bounds}
\end{equation}
These inequalities could be proven using rather standard techniques; see \cref{sec:prelim}.

To beat 2-approximation for TSP all we need to do is distinguish $\TSP \approx 2M$ from $\TSP \approx M$. In the former case, \cref{eq:K-bounds,eq:TSP-bounds} imply $K \approx M$ and in the latter we must have $K \approx M/2$. Hence, distinguishing $K \approx M$ from $K \approx M/2$ suffices for our purpose.

\subsubsection{Estimating $K$ via MST Shortening}

We define the \emph{MST shortening problem}, which is closely related to the Cut LP through its dual.
Given an MST $T$ in a metric, a shortening $T'$ is a tree with the same edges whose weights satisfy
$$
0 \leq w_{T'}(e) \leq w_T(e).
$$
That is, $T'$ is obtained from $T$ by reducing some of its edge weights.

We say a shortening $T'$ is \emph{dominated by the metric} if for any two vertices $u$ and $v$, it holds that
$$
d_{T'}(u, v) \leq d(u, v).
$$
Put differently, the distances in $T'$ are dominated by the metric distances.
We use $T^*$ to denote a dominated shortening of maximum weight.

To see how the optimal dominated shortening weight is related to $K$, consider the following intuition.
Both $K$ and $w(T^*)$ are large, that is $\approx M$, when the metric distances are essentially the same as the distances on $T$.
Considering $T^*$, observe that for any vertices $u$ and $v$, the tree distances sandwich the metric distance:
$$
d_{T^*}(u, v) \leq d(u, v) \leq d_T(u, v),
$$
where the first inequality follows from the fact that $T^*$ is dominated by the metric, and the second inequality follows from the metric property (i.e., the triangle inequality).
Therefore, $w(T^*) \approx M$ means that $d(u, v)$ is the same as $d_T(u, v)$ up to a small perturbation.

Furthermore, we remark that where $K$ lies between $M/2$ and $M$ captures how many shortcuts the metric has over the tree distances of $T$.
Consider two examples:
\begin{enumerate}
    \item Let the metric be induced by a path with unit edges. That is, the metric distance between any two vertices is equal to their distance on the path. In this case, the optimal solution of~\eqref{eq:primal} is obtained by setting $x_e = 1$ on the path. $K = n - 1 = M$.\footnote{This is more generally true whenever the metric is induced by the MST.}
    \item Let the metric be induced by a cycle with unit edges.
    In other words, taking a path of unit-weight edges as the MST, there is a shortcut of weight $1$ between its endpoints.
    In this case, the optimal solution of~\eqref{eq:primal} is obtained by setting $x_e = 1/2$ on the cycle, yielding $K = \frac{n}{2} \approx \frac{M}{2}$.
\end{enumerate}
This further relates $K$ and $w(T^*)$ as follows:
Shortcuts over the MST force a significant decrease on the tree path between the two endpoints when obtaining a dominated shortening,
whereas in the Cut LP, shortcuts allow us to decrease the values assigned to the path at the price of a (comparably small) increase in the value assigned to the shortcut.

To formalize this intuition, we prove that when $K \approx M$, it holds that $w(T^*) \approx M$. 
We further show that $w(T^*) \leq K$. This reduces the original problem to distinguishing between cases where $w(T^*) \approx M$ and $w(T^*) \lesssim M/2$.
However, there are two challenges to estimating $w(T^*)$:
(1) The edges of the MST are not known to the algorithm. In fact, computing the edges of any constant-approximate MST would require $\Omega(n^2)$ queries \cite{Indyk99,ChenKT23Steiner}.
(2) Even assuming we are given the edges of an MST, computing a dominated shortening is still non-trivial.
For example, given an MST shortening in a $(1, 2)$-metric, confirming whether it is dominated by the metric would still require $\Omega(n^2)$ queries.

\subsubsection{Dual Solutions and Steiner Trees}
Before addressing the challenges of estimating $w(T^*)$, we examine the dual of the Cut LP, and the representation of its solutions using weighted Steiner trees.

The dual LP assigns a value $y_A \geq 0$ to every cut $\delta(A)$ (which is the outgoing edges of a vertex set $A$) such that for any edge $e$, the total value assigned to the cuts containing $e$ is at most $w_e$:
\begin{equation}
K = \max\left\{\sum_{\emptyset\neq A\subsetneq V} y_A \ \middle\lvert\  \sum_{A:e\in\delta(A)} y_A \leq w_e\quad (\forall e), \quad y\geq 0\right\}. \tag{$\mathcal{D}$} \label{eq:dual}
\end{equation}
In the case of metrics, the edge set includes every pair of vertices, weighted by the metric distance between the two endpoints.

An MST shortening $T'$ naturally provides a solution for the dual LP:
For every edge $e \in T'$, let $A$ and $B$ be the vertices of the two connected components of $T' \setminus e$.
The dual solution corresponding to $T'$ assigns $y_A = w_{T'}(e)$.
Observe that this solution is feasible, meaning it satisfies the constraints of the dual LP, if the MST shortening is dominated.
To see this, note that the cuts of non-zero value $y_A$ that include an edge $(u, v)$, are exactly those corresponding to an edge on the $u$-$v$ path in $T'$.
Therefore:
$$
\sum_{A:(u, v) \in\delta(A)} y_A 
= \sum_{e \in P_{T'}(u, v)} w_{T'}(e)
= d_{T'}(u, v) \leq d(u, v).
$$

We can further generalize this to represent dual solutions using weighted Steiner trees.
That is, given a metric on vertices $V$, we consider weighted trees $F$, whose vertex set is a superset of $V$ and whose leaves are vertices of $V$.
Then, similar to MST shortenings, we can obtain a corresponding dual solution by assigning $y_A = w_e$ for every $e \in F$,
where $A$ denotes the vertices of $V$ lying in one of the components of $F \setminus e$.
This dual solution is feasible if the tree distances $d_F(u, v)$ are dominated by the metric $d(u, v)$.
It can be further argued that \emph{any} dual solution has a corresponding Steiner tree representation, using standard uncrossing techniques.


\subsection{Towards a Subquadratic Query Algorithm}

We decompose the metric into cells,
define local dual solutions in each of them,
and consider their combination, which yields a dual solution $\Fdual$ for the entire metric.
We show that $L = w(\Fdual)$ is large when $K \approx M$.
Hence, the algorithm can estimate $L$ to determine if $K \approx M$.
This approach works around computing a global MST:
By breaking down the metric into cells, it becomes easier to make inferences about small parts of the MST.

We compute a core $C$, which is a weighted Steiner tree with a subset $S$ of the metric points $V$ as its terminals, where $\card{S} = \Ot(n^{1 - a})$ for a constant $a$.
When $w(T^*) \approx M$, the core captures the \enquote{long} parts of the metric.
We use $\RC$ to denote the continuous extension of $C$,
meaning it is a metric that contains the vertices of $C$ as well as the interior points of its edges.
Each vertex $v \in V$ is mapped to a point $q_v \in \RC$.
This mapping captures the general location of the vertices in relation to each other, and is used to decompose the metric into cells.

\subsubsection{The Computed Core}
First, we compute a set of core terminals $S \subseteq V$.
Starting with a vertex set $S = \Sbase$, we add progressively finer $r$-nets to $S$, as long as doing so increases the value of $K(S)$ substantially.
Here, $\Sbase$ is a carefully chosen set that allows breaking down the MST into small subtrees and locally exploring them, the details of which are deferred to the main text.
An $r$-net refers to a vertex set $N$, such that 
\begin{enumerate}
    \item for any vertex $v \in V$, it holds that $d(v, N) \leq r$ and
    \item for any two vertices $u, v \in N$, it holds that $d(u, v) > r$.
\end{enumerate}

More precisely, we start with $r = M / n^{1 - a/10}$.
Since there exists a Hamiltonian cycle of length at most $2M$, any $r$-net $N$ satisfies $\card{N} \leq O(M / r) = O(n^{1 - a/10})$.
Furthermore, one can compute an $r$-net greedily using $O(\card{N} \cdot n)$ queries, by repeatedly adding a vertex $v$ with $d(v, N) > r$ to $N$ and querying all its distances.
The $r$-net $N$ is then added to $S$ if it holds that
\begin{equation}
K(S \cup N) \geq K(S) + \epsilon M, \label{eq:tech-SN}
\end{equation}
in which case $a$ is also replaced by $a / 10$, yielding a smaller radius $r$ for the next iteration.

It can be shown that $K(S) \leq K$ for any vertex set $S \subseteq V$.
Therefore, the process above stops within $1/\epsilon$ iterations.
At that point, either $K(S) \approx M$, which already certifies that $K$ is large and the algorithm halts,
or we have obtained a set of core terminals $S$ and an $r$-net $N$ satisfying:
\begin{equation*}
    \card{S}=\Ot(n^{1-a}),\qquad
    \card{N}=O(n^{1-a/10}),\qquad
    \text{and}
    \qquad 
    K(S\cup N)-K(S)\leq \epsilon M.
\end{equation*}

When the final $S$ and $N$ are obtained,
we compute the optimal dominated Steiner tree $C$ using $S$ as the terminals, so that $w(C) = K(S)$.
We refer to $C$ as the \emph{computed core} and to $S$ as the \emph{core terminals}.
In the analysis, we use $S$ to divide the edges in the optimal shortening $T^*$ into two groups, which we treat separately:
\begin{enumerate}
    \item \textbf{The core part, $\bm{{T^*\hull{S}}}$},
    which is the smallest subtree of $T^*$ spanning the vertices of $S$.
    \item \textbf{The hanging parts, $\bm{{T^*\setminus T^*\hull{S}}}$}, which is a collection of subtrees obtained by removing the core part from $T^*$.
\end{enumerate}
We remark that since $N$ is an $r$-net, the subtrees hanging from $T^*\hull{S \cup N}$ have depth of at most $r$.
Furthermore, due to the stopping condition \eqref{eq:tech-SN}, the discrepancy between $T^*\hull{S \cup N}$ and $T^*\hull{S}$ is small.
Therefore, the hanging parts of $T^*\hull{S}$ can essentially be thought of as subtrees of depth $r$.
This bound on the heights allows the algorithm to compute shortenings query-efficiently.

\subsubsection{Cell Decomposition}
After obtaining the core $C$, we query the distances from every vertex terminal $s \in S$ to every vertex $v \in V$, and compute a mapping $q_v \in \RC$ such that every vertex $v \in S$ is mapped to itself in $C$, and
$$
d_C(q_u, q_v) \leq d(u, v),
$$
for all vertices $u$ and $v$.
Here, $\RC$ denotes the continuous extension of $C$, that is, it includes the vertices in $C$ as well as the interior points of the edges.

The \emph{non-expansion property} of $q$ allows it to capture the locations of the vertices in the metric, relative to each other.
Formally, we prove a \emph{core alignment lemma} that states the distances induced on $S$ by $T^*$ and by $C$ are the same, up to small perturbations.
This also limits the distortion of the distances for vertices not in $S$, due to the fact that $N$ is an $r$-net.\footnote{This is, again, ignoring the discrepancy between $T^*\hull{S}$ and $T^*{\hull{S\cup N}}.$}

With the core $C$ and the mapping $q$ at hand, we can divide the metric into cells.
We call a vertex in $C$ \emph{essential} if it has degree different from $2$. This includes the leaves.
We create two kinds of cells:
(1) For each essential vertex, consider the ball of radius $R$ around it. Note that they may include parts of the edges. The connected components in the union of these balls form the \emph{buffer cells}.
(2) The remaining unassigned segments of $C$ form edge-disjoint paths. We break them down into subpaths of length $\ell$ to form the \emph{interval cells}.
We choose $R = O(M/n^{1-a/2})$ randomly, and add a random shift to the interval cells, so that the probability of any specific edge crossing between cells is bounded.

After decomposing the core $C$ into cells,
for each cell $P$, we define its vertices $V_P$ to be the set of vertices $v$ with $q_v \in P$.
Then, we define a local solution on $V_P$ (i.e., a Steiner tree) inside each cell. 
The local solutions are defined such that,
in addition to being dominated by the metric,
they can be merged into one global solution $\Fdual$.
The weight $L = w(\Fdual)$ then serves as a lower bound for $K$, and we prove it is larger than $(1 + \epsilon)\frac{M}{2}$ when $K \approx M$.

To compute the local solutions, we define \emph{regular cells}, for which the task is easier.
A cell is regular if it is an interval cell and includes at most $n^a$ vertices.\footnote{There is a third condition, which we defer to the main text.}
Given a regular cell, we compute the local solution after querying all the $O(\card{V_P}^2)$ pairs.
This takes $O(n^{1 + a})$ queries over all regular cells.
We refer to the remaining cells as the \emph{contracted cells},
which we examine in the next section.

\begin{figure}[tbp]
    \centering
    \begin{tikzpicture}[
    x=0.78cm,y=0.78cm,font=\small,
    core/.style={draw=black!75,line width=0.9pt,line cap=round,line join=round},
    buffer fill/.style={fill=blue!4,draw=none},
    buffer outline/.style={draw=blue!40!gray,line width=0.85pt,line cap=round},
    interval/.style={draw=orange!60!black,fill=orange!9,line width=0.65pt,rounded corners=0.7pt},
    route/.style={draw=red!70!black,line width=1.65pt,line cap=round,line join=round},
    essential/.style={circle,fill=black,inner sep=1.85pt},
    port/.style={circle,draw=black!70,fill=white,line width=0.65pt,inner sep=1.6pt},
    label point/.style={circle,fill=red!70!black,inner sep=1.8pt},
    annotation/.style={font=\normalsize,text=black,inner sep=0.5pt}
]
    \draw[interval,rotate around={38.4537:(1.6666,-0.6530)}] (1.3105,-0.8630) rectangle (2.0228,-0.4430);
    \draw[interval,rotate around={38.4537:(2.3000,-0.1500)}] (1.8473,-0.3600) rectangle (2.7527,0.0600);
    \draw[interval,rotate around={38.4537:(2.9334,0.3530)}] (2.5772,0.1430) rectangle (3.2895,0.5630);
    \draw[interval,rotate around={-35.2176:(1.6791,2.8383)}] (1.3780,2.6283) rectangle (1.9802,3.0483);
    \draw[interval,rotate around={-35.2176:(2.3000,2.4000)}] (1.8411,2.1900) rectangle (2.7589,2.6100);
    \draw[interval,rotate around={-35.2176:(2.9209,1.9617)}] (2.6198,1.7517) rectangle (3.2220,2.1717);
    \draw[interval,rotate around={-112.0679:(5.0315,3.7444)}] (4.6894,3.5344) rectangle (5.3737,3.9544);
    \draw[interval,rotate around={-112.0679:(4.7500,3.0500)}] (4.3429,2.8400) rectangle (5.1571,3.2600);
    \draw[interval,rotate around={-112.0679:(4.4685,2.3556)}] (4.1263,2.1456) rectangle (4.8106,2.5656);
    \draw[interval,rotate around={-19.2307:(5.3565,0.7268)}] (5.0040,0.5168) rectangle (5.7090,0.9368);
    \draw[interval,rotate around={-19.2307:(6.1500,0.4500)}] (5.6621,0.2400) rectangle (6.6379,0.6600);
    \draw[interval,rotate around={-19.2307:(6.9435,0.1732)}] (6.5910,-0.0368) rectangle (7.2960,0.3832);
    \draw[interval,rotate around={-106.5571:(7.9549,-1.4607)}] (7.6381,-1.6707) rectangle (8.2718,-1.2507);
    \draw[interval,rotate around={-106.5571:(7.7500,-2.1500)}] (7.3477,-2.3600) rectangle (8.1523,-1.9400);
    \draw[interval,rotate around={-106.5571:(7.5451,-2.8393)}] (7.2282,-3.0493) rectangle (7.8619,-2.6293);
    \draw[interval,rotate around={36.4692:(10.5072,1.1445)}] (10.2691,0.9345) rectangle (10.7453,1.3545);
    \draw[interval,rotate around={36.4692:(11.0658,1.5573)}] (10.6093,1.3473) rectangle (11.5223,1.7673);
    \draw[interval,rotate around={36.4692:(11.8000,2.1000)}] (11.3435,1.8900) rectangle (12.2565,2.3100);
    \draw[interval,rotate around={36.4692:(12.5342,2.6427)}] (12.0777,2.4327) rectangle (12.9907,2.8527);
    \draw[interval,rotate around={36.4692:(13.0928,3.0555)}] (12.8547,2.8455) rectangle (13.3309,3.2655);
    \draw[interval,rotate around={-30.2564:(10.7450,-0.3262)}] (10.3447,-0.5362) rectangle (11.1452,-0.1162);
    \draw[interval,rotate around={-30.2564:(11.4953,-0.7639)}] (11.0268,-0.9739) rectangle (11.9638,-0.5539);
    \draw[interval,rotate around={-30.2564:(12.3047,-1.2361)}] (11.8362,-1.4461) rectangle (12.7732,-1.0261);
    \draw[interval,rotate around={-30.2564:(13.0550,-1.6738)}] (12.6548,-1.8838) rectangle (13.4553,-1.4638);

    \path[buffer fill] (4.0000,1.2000) ellipse[x radius=1.1200,y radius=0.8800];
    \path[buffer fill] (8.3000,-0.3000) ellipse[x radius=1.1200,y radius=0.8800];
    \path[buffer fill] (9.5000,0.4000) ellipse[x radius=1.1200,y radius=0.8800];
    \path[buffer fill] (0.6000,-1.5000) ellipse[x radius=1.1200,y radius=0.8800];
    \path[buffer fill] (0.6000,3.6000) ellipse[x radius=1.1200,y radius=0.8800];
    \path[buffer fill] (5.5000,4.9000) ellipse[x radius=1.1200,y radius=0.8800];
    \path[buffer fill] (7.2000,-4.0000) ellipse[x radius=1.1200,y radius=0.8800];
    \path[buffer fill] (14.1000,3.8000) ellipse[x radius=1.1200,y radius=0.8800];
    \path[buffer fill] (14.3000,-2.4000) ellipse[x radius=1.1200,y radius=0.8800];

    \draw[buffer outline] (4.0000,1.2000) ellipse[x radius=1.1200,y radius=0.8800];
    \draw[buffer outline] (8.4027,0.5763) arc[start angle=84.7386,end angle=348.4436,x radius=1.1200,y radius=0.8800];
    \draw[buffer outline] (9.3973,-0.4763) arc[start angle=264.7386,end angle=528.4436,x radius=1.1200,y radius=0.8800];
    \draw[buffer outline] (0.6000,-1.5000) ellipse[x radius=1.1200,y radius=0.8800];
    \draw[buffer outline] (0.6000,3.6000) ellipse[x radius=1.1200,y radius=0.8800];
    \draw[buffer outline] (5.5000,4.9000) ellipse[x radius=1.1200,y radius=0.8800];
    \draw[buffer outline] (7.2000,-4.0000) ellipse[x radius=1.1200,y radius=0.8800];
    \draw[buffer outline] (14.1000,3.8000) ellipse[x radius=1.1200,y radius=0.8800];
    \draw[buffer outline] (14.3000,-2.4000) ellipse[x radius=1.1200,y radius=0.8800];

    \draw[core] (0.6000,-1.5000)--(4.0000,1.2000);
    \draw[core] (0.6000,3.6000)--(4.0000,1.2000);
    \draw[core] (5.5000,4.9000)--(4.0000,1.2000);
    \draw[core] (4.0000,1.2000)--(8.3000,-0.3000);
    \draw[core] (8.3000,-0.3000)--(9.5000,0.4000);
    \draw[core] (8.3000,-0.3000)--(7.2000,-4.0000);
    \draw[core] (9.5000,0.4000)--(14.1000,3.8000);
    \draw[core] (9.5000,0.4000)--(14.3000,-2.4000);

    \draw[route] (2.9334,0.3530)--(4.0000,1.2000)--(8.3000,-0.3000)--(9.5000,0.4000)--(11.0658,1.5573);
    \node[essential] at (4.0000,1.2000) {};
    \node[essential] at (8.3000,-0.3000) {};
    \node[essential] at (9.5000,0.4000) {};
    \node[essential] at (0.6000,-1.5000) {};
    \node[essential] at (0.6000,3.6000) {};
    \node[essential] at (5.5000,4.9000) {};
    \node[essential] at (7.2000,-4.0000) {};
    \node[essential] at (14.1000,3.8000) {};
    \node[essential] at (14.3000,-2.4000) {};
    \node[port] at (3.2123,0.5744) {};
    \node[port] at (5.0236,0.8429) {};
    \node[port] at (5.6894,0.6107) {};
    \node[port] at (6.6106,0.2893) {};
    \node[port] at (7.2764,0.0571) {};
    \node[port] at (10.3158,1.0030) {};
    \node[port] at (10.6987,1.2860) {};
    \node[label point] at (2.9334,0.3530) {};
    \node[label point] at (11.0658,1.5573) {};
    \node[annotation] at (3.3323,1.0544) {$p_1$};
    \node[annotation] at (4.8036,1.2629) {$p_2$};
    \node[annotation] at (5.5694,0.1607) {$p_3$};
    \node[annotation] at (10.9587,0.8660) {$p_k$};
    \node[annotation] at (3.1934,-0.0770) {$q_u$};
    \node[annotation] at (10.8058,1.9873) {$q_v$};
    \node[annotation] at (-0.7500,1.0000) {$C$};
\end{tikzpicture}
    \caption{Cell decomposition of the core. The oval shapes represent the buffer cells, and the rectangular shapes represent the interval cells.}
    \label{fig:core-cell-decomposition}
\end{figure}
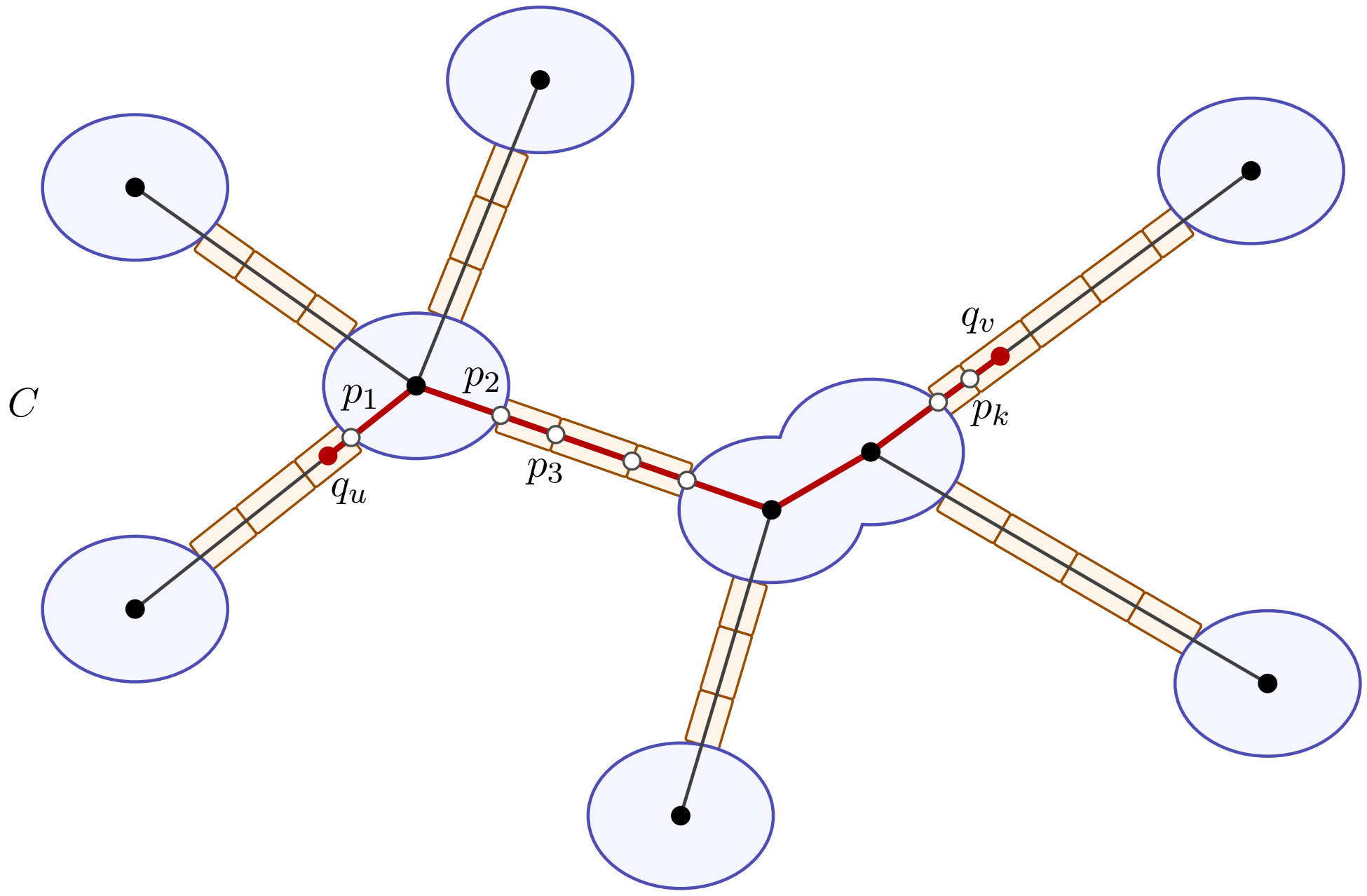

\subsubsection{Contracted Cells and Conflict Graphs}
For contracted cells, the desired local solutions cannot be explicitly computed, partly due to the large number of vertices they may include.
Therefore, the algorithm estimates their size using sampling and local estimations.

On a high level, we consider fragments of the hanging parts of the MST, $T\setminus T\hull{S}$, that lie within the cell.
The fragments are defined such that the algorithm can locally explore and recover them.
For each MST fragment, we define a dominated shortening (with some additional properties).
The total weight of these fragment shortenings is estimated using \emph{importance sampling}.
Then, to obtain the local solution, we combine the fragments by contracting all their roots together.
Combining the solutions this way may result in pairs that violate the domination constraint.
Therefore, we need to further shorten some of the edges to ensure we get a feasible dual solution.
The rest of this section demonstrates how this combination is obtained through the minimum vertex cover of an auxiliary \emph{conflict graph}.


Let $T_1$ and $T_2$ be two subtrees sharing a root $r$ and otherwise disjoint.
Suppose that each subtree has been shortened such that it is dominated by the metric.
Formally, for all $i \in \{1, 2\}$ and $u, v \in T_i$, it holds that
$$
d_{T_i}(u, v) \leq d(u, v).
$$
Our goal is to combine these subtrees into a single tree $T_\cup = T_1 \cup T_2$ rooted at $r$ that is also dominated by the metric.

Observe that, while each tree is dominated individually, the same need not be true about $T_1 \cup T_2$.
That is, there may exist vertices $u \in T_1$ and $v \in T_2$ such that
$$
d_{T_\cup}(u, v) > d(u, v).
$$
Indeed, although we are guaranteed that
$$d_{T_\cup}(u, v) = d_{T_1}(u, r) + d_{T_2}(r, v) \leq d(u, r) + d(r, v), $$
still $d(u, v)$ may be significantly smaller than $d(u, r) + d(r, v)$.
Thus, further shortening of the trees is necessary to ensure that $T_{\cup}$ is dominated.

\begin{figure}
\centering
\begin{tikzpicture}[x=0.95cm,y=0.95cm,
    branch/.style={draw=black!75,line width=0.85pt,line cap=round,line join=round},
    root path/.style={draw=black!75,line width=1.25pt,line cap=round,line join=round},
    metric/.style={draw=black,line width=0.85pt,line cap=round,line join=round,dash pattern=on 3pt off 2pt},
    conflict/.style={draw=black!65,line width=0.85pt,line cap=round,line join=round},
    zero/.style={draw=black!25,line width=1.0pt,line cap=round,line join=round}
]
    \path[draw=blue!40!gray,fill=blue!4,line width=0.6pt] (0.000,4.500) .. controls (-1.300,4.150) and (-2.600,2.400) .. (-2.600,0.200) .. controls (-2.600,-0.750) and (-0.800,-0.800) .. (-0.550,0.150) .. controls (-0.300,2.500) and (-0.200,3.800) .. (0.000,4.500) -- cycle;
    \path[draw=blue!40!gray,fill=blue!4,line width=0.6pt] (0.000,4.500) .. controls (1.300,4.150) and (2.600,2.400) .. (2.600,0.200) .. controls (2.600,-0.750) and (0.800,-0.800) .. (0.550,0.150) .. controls (0.300,2.500) and (0.200,3.800) .. (0.000,4.500) -- cycle;
    \draw[root path] (0.000,4.500) .. controls (-0.850,3.300) and (-1.550,1.800) .. (-1.500,0.150);
    \draw[root path] (0.000,4.500) .. controls (0.850,3.300) and (1.550,1.800) .. (1.500,0.150);
    \draw[metric] (-1.500,0.150) .. controls (-1.150,-1.100) and (1.150,-1.100) .. (1.500,0.150);
    \draw[draw=black,fill=black,line width=0.6pt] (0.000,4.500) circle[radius=1.65pt];
    \draw[draw=black,fill=black,line width=0.6pt] (-1.500,0.150) circle[radius=1.65pt];
    \draw[draw=black,fill=black,line width=0.6pt] (1.500,0.150) circle[radius=1.65pt];
    \node[font=\small,text=black,inner sep=0pt] at (0.000,4.830) {$r$};
    \node[font=\small,text=black,inner sep=0pt] at (-1.790,0.100) {$u$};
    \node[font=\small,text=black,inner sep=0pt] at (1.790,0.100) {$v$};
    \node[font=\small,text=black,inner sep=0pt] at (-2.350,3.800) {$T_1$};
    \node[font=\small,text=black,inner sep=0pt] at (2.350,3.800) {$T_2$};
    \node[font=\small,text=black,inner sep=0pt] at (-3.000,2.500) {$d_{T_1}(u,r)$};
    \node[font=\small,text=black,inner sep=0pt] at (3.000,2.500) {$d_{T_2}(v,r)$};
    \node[font=\small,text=black,inner sep=0pt] at (0.000,-1.120) {$d(u,v)$};
\end{tikzpicture}
\caption{
    Two dominated subtrees sharing a root.
    For $u \in T_1$ and $v \in T_2$, the distance through the root,
    $d_{T_1}(u,r)+d_{T_2}(v,r)$, may exceed the direct metric distance $d(u,v)$.
}
\label{fig:merge-subtrees-setup}
\end{figure}
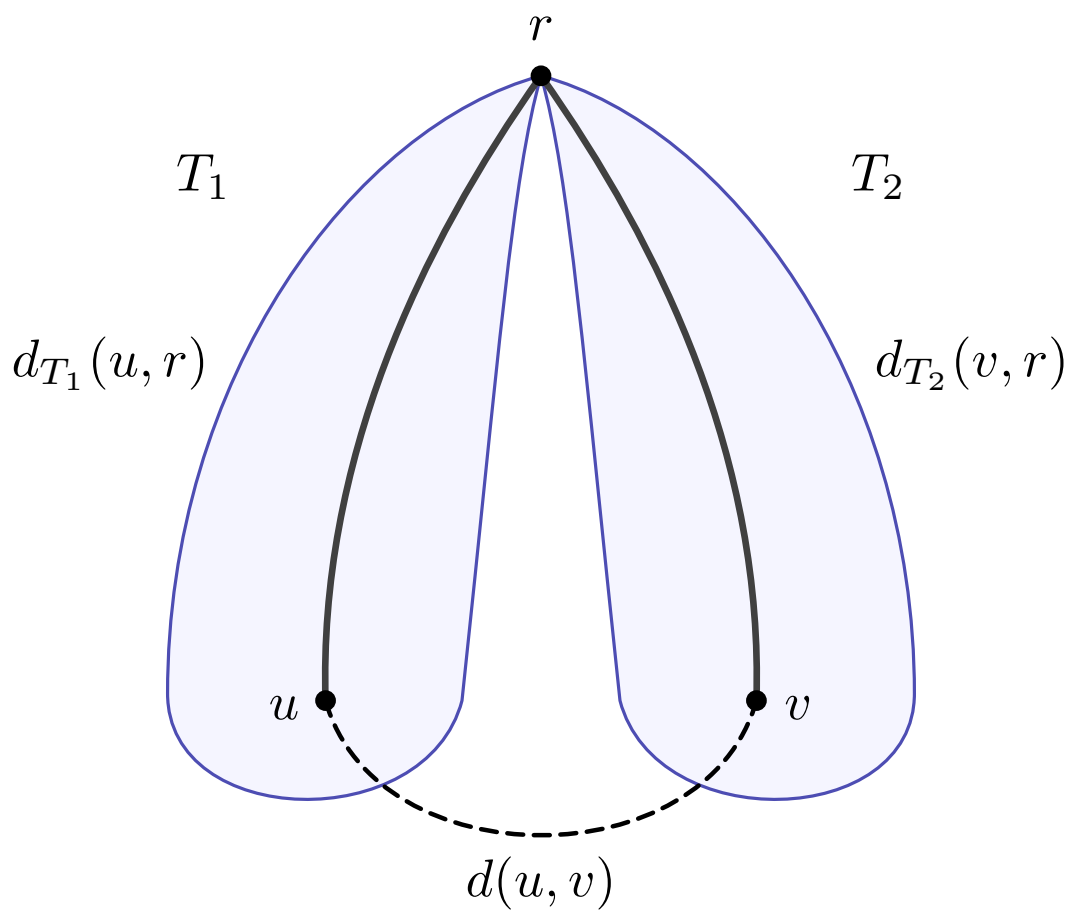

We begin by subdividing each tree edge into a path of segments.
For simplicity, assume that all edge weights are integral, and replace an edge of weight $w$ with a path of $w$ unit-weight segments.
We only consider integral shortenings obtained by \emph{erasing} segments, i.e., setting their weights to zero.
Later, we argue that a similar discretization into integer multiples of $\sigma = O(\epsilon M/n)$ is a sufficiently fine scale for obtaining an optimal shortening.
We further restrict ourselves to shortenings in which the erased segments form a rooted subtree.
This is without loss of generality, as any erased segment can be moved towards the root without violating any conditions.

\begin{figure}
\centering
\input{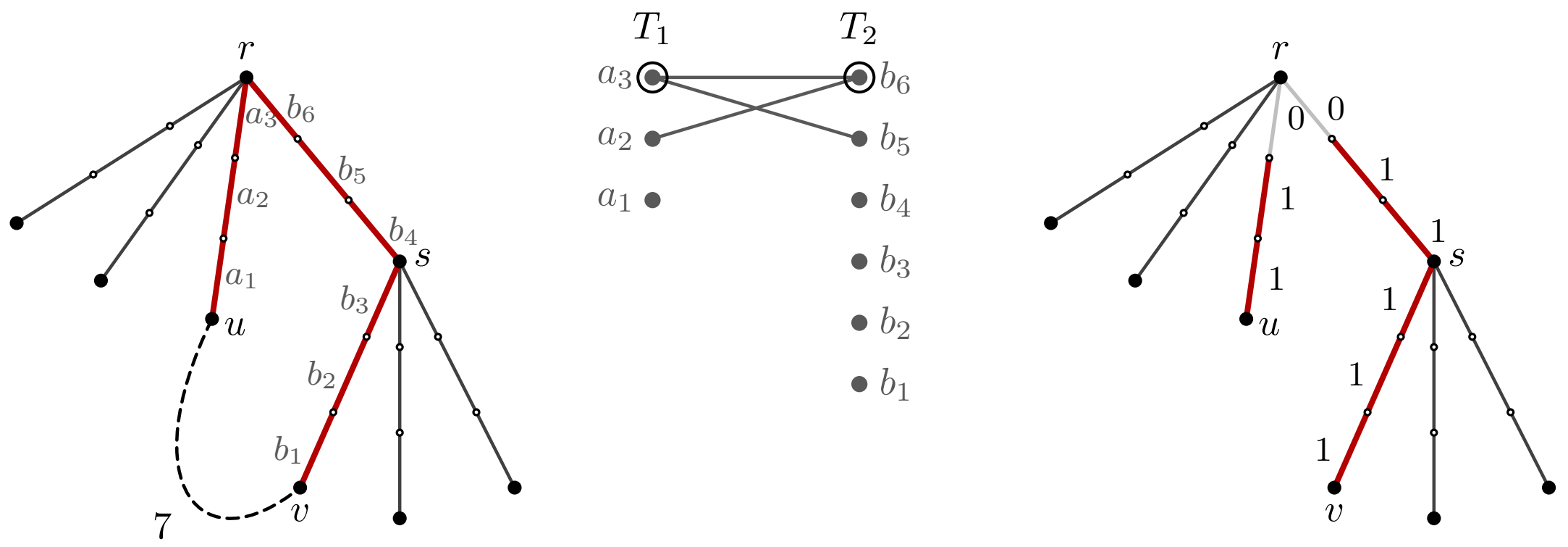}
\caption{
    Consider the metric induced by the edges displayed in (a), and let the initial weights be $3$ for all tree edges.
    In the conflict graph (b), segments above $u$ and $v$ have a conflict for $(i, j) = (2, 6), (3,6), (3, 5)$, that is, when $i + j > 7$.
}
\label{fig:conflict-graph}
\end{figure}

To construct the shortening, we introduce an auxiliary \emph{conflict graph} whose vertices represent the segments from the two subtrees.
For vertices $u \in T_1$ and $v \in T_2$, let $h_u, h_v$ denote their distance to the root.
Consider the $i$-th segment $e_1$ on the path from $u$ to $r$ and the $j$-th segment $e_2$ on the path from $v$ to $r$,
where $i \leq h_u$ and $j \leq h_v$.
We say that $e_1$ and $e_2$ are in conflict if
$$
i + j > d(u, v),
$$
and represent this conflict using an edge between $e_1$ and $e_2$ in the auxiliary graph. We note that the same pair of segments may have a conflict due to multiple pairs of vertices in their respective subtrees (see \cref{fig:conflict-graph}).

The conflict between $e_1$ and $e_2$ reflects the following fact:
at least one of $e_1$ and $e_2$ must be erased to achieve domination by the metric, since if neither is erased, then $d_{T_\cup}(u, r)$ and $d_{T_\cup}(v, r)$ would be at least $i$ and $j$, respectively.
It follows that a shortening yields a dominated tree if and only if the erased segments form a vertex cover in the conflict graph.
Hence, finding the optimal (integral) shortening is reduced to computing a minimum vertex cover in the conflict graph.
We remark that an algorithm with subquadratically many queries does not have direct access to either the conflict graph or the MST itself. Overcoming this obstacle introduces additional challenges, which we address in the main text.

\section{Preliminaries}\label{sec:prelim}

\paragraph{Problem Definition.}
In the \emph{metric traveling salesperson problem} (metric TSP),
we are given a metric $(V, d)$, meaning $d: V\times V \to \mathbb{R}_{\geq 0}$ satisfies 
$$
d(u, v) \leq d(u, w) + d(w, v),
$$
for all vertices $u, v, w \in V$.
The goal is to estimate the length of the shortest Hamiltonian cycle, which is a cycle that visits every vertex exactly once.

\paragraph{Query Model.}
We can access the metric through entry queries, meaning that for a vertex pair $(u, v)$, we can query $d(u, v)$.
In the paper, we use \emph{row query} at a vertex $v$ to refer to querying the distance between $v$ and every other vertex. This has a query complexity of $n - 1$.

\paragraph{Notations and Definitions.}
We use $n = \card{V}$ to denote the number of vertices, $T$ to denote the minimum spanning tree, $M = w(T)$ to denote its weight, $K$ to denote the optimal value of the Cut LP, and $\TSP$ to denote the weight of the shortest Hamiltonian cycle.
We let $D = M - K$.
We use $\Ot(\cdot)$ to suppress logarithmic factors in $n$.
For a value $X$, we use $\widehat{X}$ to denote its estimation.
Throughout the paper, we assume access to an estimation $\hM$ satisfying:
$$
M \leq \hM \leq (1 +O (\epsilon)) M,
$$
which is obtainable using $\Ot(n/\epsilon^{O(1)})$ queries \cite{CzumajS04}.
We also assume distinct edge weights by tie-breaking through edge IDs.

Given a tree $T$, we use subtree to refer to any connected subgraph.
To refer to the subtree which includes all the descendants of its root, we use \emph{hanging subtree}.
Given a subset of vertices $S \subseteq V(T)$, we use $T\hull{S}$ to denote the minimal subtree spanning $S$.

Given a metric on $V$, and a set of roots $S \subseteq V$, we define the
\emph{rooted minimum spanning forest} (rooted MSF), as a forest with minimum weight, such that each component includes exactly one root from $S$.

Given a tree $T$, it induces a metric on its vertices,
where for $u, v \in V(T)$, the distance $d_T(u, v)$ is equal to the weight of the path from $u$ to $v$ in $T$.
We use $\RT$ to denote the continuous extension of this tree metric.
The points in $\RT$ consist of the vertices of $T$ plus the interior points of all its edges.
That is, for edge $e = (u, v)$ and $\alpha \in (0, 1)$, there is an interior point $p$ on $e$ that satisfies
$$
d_{\RT}(u, p) = \alpha w_T(e)
\qquad
\text{and}
\qquad
d_{\RT}(v, p) = (1 - \alpha) w_T(e).
$$
We use $d_{\RT}(\cdot,\cdot)$ and $d_T(\cdot, \cdot)$ interchangeably.

We remark that the subtrees in $\RT$,\footnote{which is a closed connected subset of $\RT$} which may include parts of the edges, have a Helly number $2$.
That is,
\begin{lemma}\label{lem:tree-helly}
    Any finite collection of pairwise-intersecting subtrees $\mc{F}$ 
    has a common point, which is a member of every subtree.
\end{lemma}
\begin{proof}
    Root the tree at an arbitrary point $r \in \RT$.
    For each subtree $F \in \mc{F}$, let $a_F$ be its closest point to the root.
    Observe that any path from $r$ to a point in $F$ goes through $a_F$.

    Let $F$ be a subtree such that $a_F$ is furthest from $r$.
    We claim that $a_F$ belongs to every subtree.
    Take any $F' \in \mc{F}$ and a point $x \in F \cap F'$.
    The path from $r$ to $x$ goes through both $a_F$ and $a_{F'}$.
    By the choice of $F$, we have that $a_F$ lies on the path from $a_{F'}$ to $x$, and therefore $a_F \in F'$.
\end{proof}

Next, we include some folklore properties of the Cut LP.
\begin{claim} \label{clm:K-bounds}
It holds that
$$
\frac{M}{2} \leq K \leq M.
$$
\end{claim}
\begin{proof}
The upper bound follows from the fact that the LP is a relaxation of the MST problem, i.e., letting $x_e = 1$ for the MST edges yields a feasible solution to \eqref{eq:primal}.
For the lower bound, let $w_1 < w_2 < \ldots < w_N$ denote the distinct weights in the metric, let $w_0 = 0$, and let $C_i$ denote the connected components w.r.t.\ edges of weight at most $w_{i-1}$.
Then, it holds that 
$$M = \sum_i (w_i - w_{i-1})(\card{C_i} - 1).$$
Now, let $E_i$ denote the edges of weight at least $w_i$. Then, it holds that:
$$
K = \sum_e w_ex_e 
= \sum_i (w_i - w_{i-1}) x(E_i)
\overset{(*)}{\geq} \sum_i (w_i - w_{i-1}) \frac{\card{C_i} - 1}{2} \geq \frac{M}{2},
$$
where $(*)$ can be obtained by summing  the LP condition $x(\delta(S)) \geq 1$ for $S \in C_{i}$.
\end{proof}

\begin{claim} \label{clm:TSP-bounds}
It holds that
$$
\max(M, 2K) \leq \TSP \leq M + K.
$$
\end{claim}
\begin{proof}
    For the lower bound, note that a Hamiltonian cycle traverses any cut at least twice. Therefore, setting $x_e = \frac{1}{2}$ on the cycle yields a feasible solution of \eqref{eq:primal}, and $K \leq \frac{1}{2}\TSP$.
    Also, removing any edge from the Hamiltonian cycle yields a spanning tree (which is a path).
    Therefore, $\TSP \geq M$.

    For the upper bound, let $O$ be the set of odd-degree vertices in $T$:
    $$
    O = \{v \mid \deg_T(v) \text{ is odd}\}.
    $$
    Call an edge set $F$ an $O$-join\footnote{These are usually referred to as $T$-joins. We use $O$-join to avoid confusion with the minimum spanning tree $T$.} if it satisfies
    \[
    \deg_F(v)\equiv
    \begin{cases}
    1 \pmod 2, & v\in O,\\
    0 \pmod 2, & v\notin O.
    \end{cases}
    \]

    It can be shown that the $O$-join polytope, which is the convex hull of the vectors corresponding to $O$-joins, is dominated by the following \cite{EdmondsJ73}:
    \begin{equation}
        \left\{
            x \geq 0 \mid
            x(\delta(S)) \geq 1, \quad \forall S \subseteq V \text{ with } |S \cap O| \text{ odd} \right\}.
            \label{eq:t-join-poly}
    \end{equation}
    That is, for any set of non-negative weights $w$ and any feasible solution $x$ for \eqref{eq:t-join-poly},
    there an $O$-join of weight at most $\sum_e w_ex_e$.

    The optimal solution $x$ for the primal LP \eqref{eq:primal} is a feasible solution for \eqref{eq:t-join-poly} above, since it satisfies the condition $x(\delta(S)) \geq 1$ for any $S$.
    Therefore, there exists an $O$-join $F$ of weight at most $K$.

    Finally, a Hamiltonian cycle of weight at most $M + K$ can be obtained by taking the multiset union of the edges of $F$ and $T$.
    This yields an Eulerian multigraph where every vertex has an even degree.
    Hence, the edges can be cyclically ordered such that each edge is adjacent to the next.
    Due to the triangle inequality, and since $T$ spans all the vertices, by deleting repeated vertices, this can be transformed into a cycle that visits every vertex exactly once and has a weight of at most $M + K$.
\end{proof}

Finally, we include a well-known technique called \emph{importance sampling}, which follows from
Bernstein's inequality; see, e.g., \cite[Section~2.8]{BLM13}.

\begin{lemma}[Importance Sampling]\label{lem:importance-sampling}
Let $x_1,\ldots,x_n$ be unknown nonnegative values that can be evaluated individually, and let $b_1,\ldots,b_n>0$ be known bounds satisfying $x_i\leq b_i$.
Write $X=\sum_i x_i$ and $B=\sum_i b_i$, and let $U>0$ be a known upper bound on $X$.

Sample $s$ indices $I_1,\ldots,I_s$ independently, with
$\Pr[I_j=i]=p_i=b_i/B$, and define the unbiased estimator
\begin{equation*}
    \widehat X
    =\frac{1}{s}\sum_{j=1}^s\frac{x_{I_j}}{p_{I_j}}.
\end{equation*}
For any $\epsilon,\delta\in(0,1)$, taking
\begin{equation*}
    s\geq
    \left\lceil
        \frac{3B}{\epsilon^2 U}\ln\frac{2}{\delta}
    \right\rceil
\end{equation*}
ensures that
\begin{equation*}
    \Pr\!\left[
        |\widehat X-X|>\epsilon U
    \right]\leq\delta.
\end{equation*}
\end{lemma}
\section{The Dual LP and MST Shortenings}
In this section, we present the main tools for estimating $K$.
Mainly, we show that optimizing the dual LP is equivalent to computing Steiner trees dominated by the metric,
and then prove that when $K \approx M$, the problem is equivalent to computing the maximum weight MST shortenings dominated by the metric.

\subsection{Dual LP and Tree Representation}
We restate the dual LP here:
\begin{equation}
K = \max\left\{\sum_{\emptyset\neq A\subsetneq V} y_A \ \middle\lvert\  \sum_{A:e\in\delta(A)} y_A \leq w_e\quad (\forall e), \quad y\geq 0\right\}. 
\end{equation}
Here $e = (u, v)$ ranges over all pairs $u \neq v \in V$,
and $w_e$ is given by the metric distance $d(u, v)$.

We handle dual solutions by representing them using weighted Steiner trees whose vertex set contains $V$, and whose leaves are members of $V$.
Such a tree $F$ naturally corresponds to a dual solution:
Start with $y_A = 0$ for all $A$.
For any edge $e \in F$, let $A$ denote the vertices of $V$ lying in one of the two connected components of $F \setminus e$.
Considering edge $e$, add $w_F(e)$ to $y_A$.

We say a Steiner tree $F$ spanning $V$ is \emph{dominated by the metric} if it satisfies:
$$
d_F(u, v) \leq d(u, v),
$$
where $d_F(u, v)$ denotes the sum of edge weights on the path from $u$ to $v$ in $F$.
In other words, the lengths of the paths are dominated by the metric distances.

\begin{claim}
    The corresponding dual solution $y$ of a tree $F$ satisfies the dual constraints if and only if the tree is dominated by the metric.
\end{claim}
\begin{proof}
Observe that given a vertex pair $u, v \in V$, the cuts with positive $y$-values that separate $u$ and $v$ are exactly those corresponding to the edges of the $u$-$v$ path in $F$.
Therefore, it holds that:
$$
\sum_{A:(u, v)\in\delta(A)} y_A = \sum_{e \in P_F(u, v)}w_F(e) = d_F(u, v),
$$
which concludes the proof.
\end{proof}

Finally, considering Steiner trees suffices for optimizing the dual LP.

\begin{lemma} \label{lem:dual-tree-rep}
    For any feasible dual solution $y$,
    there exists a weighted tree $F$ that is dominated by the metric,
    and satisfies
    $$
    w(F) = \sum_A y_A.
    $$
\end{lemma}
\begin{proof}
    Fix any vertex $x$.
    We can assume without loss of generality that dual solutions only assign non-zero values to $y_A$ with $x \notin A$. Otherwise, the value of $y_A$ can be transferred to $y_{V \setminus A}$.
    
    Take a feasible dual solution $y$.
    We show that there exists a dual solution $y'$ with $\card{y'} = \card{y}$,
    such that the sets $A$ with $y'_A > 0$ form a laminar family (i.e., a hierarchy).
    That is, for any $A, B \subseteq V$ with $y'_A, y'_B > 0$, one of the following holds:
    $$
    A \cap B = \emptyset \qquad \text{or} \qquad A \subseteq B \qquad \text{or} \qquad B \subseteq A.
    $$

    This is proven through standard uncrossing techniques.
    Let $y'$ be a feasible dual solution with $\card{y'} = \card{y}$ such that it minimizes:
    $$
    \Phi(y') = \sum_A y'_A \cdot \card{A}(n - \card{A}).
    $$
    Assume for the sake of contradiction that the support of $y'$ does not form a laminar family.
    That is, there are sets $A$ and $B$ with positive $y'$-values such that $A \cap B \neq \emptyset$, and neither is contained in the other.
    Let $\delta = \min(y'_A, y'_B)$.
    We obtain another dual solution $y''$ by decreasing $y'_A$ and $y'_B$ by $\delta$, and increasing $y'_{A \cup B}$ and $y'_{A \cap B}$ by $\delta$ (note that $A \cup B$ is a valid cut since $x \notin A \cup B$).
    It can be seen that $y''$ is feasible, $\card{y''} = \card{y'}$,  and $\Phi(y'') < \Phi(y')$ which is a contradiction.

    Given the feasible dual solution $y'$, with $\card{y'} = \card{y}$, whose support forms a laminar family, we can construct a Steiner tree $F$ as follows.
    Create 
    a node representing $V$,
    a node representing $A$ for each $A$ with $y'_A > 0$,
    and a node representing $A = \{v\}$ for $v \in V$.
    Let the node representing $V$ be the root.
    For any other node $A$, let its parent in the tree be $B$, the smallest set among the created nodes that contains $A$.
    Finally, let $w_F(A, B) = y'_A$.
    It can be seen that for any $(u, v)$, any cut $A$ separating $u$ and $v$ contributes $y'_A$ to the $u$-$v$ path in $F$.
    Therefore, $y'$ is the corresponding dual solution of $F$, which concludes the proof.
\end{proof}

To determine whether $K$ is large, the algorithm explores a particular dominated Steiner tree $\Fdual$ and evaluates its size.
$\Fdual$ is constructed by combining local solutions in each cell.
We formally define $\Fdual$ in the later sections.
Our approach has two main components:
\begin{enumerate}
    \item Establishing that $w(\Fdual)$ is large when $M - K \leq \epsilon M$, and
    \item Developing an algorithm to approximate $w(\Fdual)$.
\end{enumerate} 
We start by considering MST shortenings, which serve as the key motivation for the definition of $\Fdual$.

\subsection{MST Shortening}
A key observation in our approach is that when $D = M - K \leq \epsilon M$, the MST largely captures the structure of the metric.
As such, to optimize the dual program \eqref{eq:dual}, we can restrict our attention to a subclass of dominated trees, \emph{MST shortenings}.
That is, to decrease the weights of the edges on the MST such that the resulting tree corresponds to a feasible solution of \eqref{eq:dual}.
When $D$ is small, there exists such a dominated MST shortening that is nearly optimal:

\begin{lemma} \label{lem:mst-shortening}
    There exists a dominated shortening $T_0$ of $T$, such that 
    $$
    w(T_0) \geq K - D = M - 2D.
    $$
\end{lemma}
\begin{proof}

Consider the MST $T$ and an optimal dominated tree $F$.
Map every MST edge $e = (u, v)$, to the $u$-$v$ path on $F$, and charge $w_F(e')$ to each edge $e' \in F$ on this path.
Let $W$ be the sum of the path lengths on $F$, that is, the total charge.
Since $F$ is dominated by the metric, each edge $e \in T$ is mapped to a path $P$ with $w_F(P) \leq w_T(e)$.
Therefore, it holds that $$
W \leq M.
$$

Since $T$ spans all the vertices $V$ and the leaves of $F$ are all members of $V$, the paths cover $F$.
Let $E_1$ be the edges of $F$ that have been charged exactly once, and let $E_2$ denote the remainder of the edges, which are edges where there is an overlap of the paths.
Then, it holds that:
$$
W \geq w_F(E_1) + 2w_F(E_2) = w(F) + w_F(E_2),
$$
and rearranging gives
$$
w_F(E_2) \leq W - w(F) \leq M - K = D.
$$

Now, to construct the MST shortening, we first modify $F$.
Set the weights on the overlap edges $E_2$ to zero, obtaining a new Steiner tree $F'$.
We construct a shortening $T_0$, by letting any edge $e = (u, v) \in T$ have weight $d_{F'}(u, v)$.
This is a shortening of $T$ since
$$
w_{T_0}(e) = d_{F'}(u, v) \leq d_F(u, v) \leq w_T(e),
$$
where the first inequality follows from the fact that $F'$ was obtained by setting weights of $F$ to zero, and the second inequality holds because $F$ is dominated by the metric.
Furthermore, since the overlap edges have weight $0$ in $F'$, it holds that
$$
w(T_0) = w(F') = w(F) - w_F(E_2) \geq K - D.
$$

It remains to show that $T_0$ is dominated by the metric.
Take any vertex pair $(u, v)$.
Let $e_1, e_2, \ldots e_k \in T$ denote the edges of the $u$-$v$ path on $T$.
For each $e_i = (x, y)$, consider the $x-y$ path on $F'$, and append them to get a sequence of edges $P = e'_1, e'_2, \ldots, e'_K$.
This sequence forms a walk on $F'$ that starts at $u$ and ends at $v$.
By definition, $w_{T_0}(e_i)$ equals the length of its corresponding path in $F'$, which is equal to the weight of the same path on $F$ with the overlap edges ignored.
Observe that any edge in $P$ that is off the $u$-$v$ path must be an overlap edge.
Therefore, it holds that
\begin{equation*}
d_{T_0}(u, v) = \sum_i w_{T_0}(e_i) = d_{F'}(u, v) \leq d_F(u, v) \leq d(u, v). \qedhere
\end{equation*}
\end{proof}

This implies the following useful relation between $K(S)$ and $T\hull{S}$,
where recall that $K(S)$ is the optimal value of the Cut LP on the metric induced by a vertex set $S$,
and $T\hull{S}$ is the minimal subtree of $T$ spanning $S$.

\begin{lemma}\label{lem:k-vs-hull}
    It holds that 
    $$
    w(T\hull{S}) - 2D \leq K(S) \leq w(T\hull{S}).
    $$
\end{lemma}
\begin{proof}
    For the upper bound, note that since $T\hull{S}$ spans $S$, we can construct a Hamiltonian cycle of $S$ with length at most $2w(T\hull{S})$. Setting $x_e = 1/2$ on the edges of this cycle is a feasible solution of the Cut LP \eqref{eq:primal}.

    For the lower bound, take an MST shortening $T_0$ with $w(T_0) \geq M - 2D$.
    Such a $T_0$ exists due to \cref{lem:mst-shortening}.
    Since $T_0$ is dominated by the metric, $T_0\hull{S}$ is dominated by the metric induced on $S$.
    That is, $T_0\hull{S}$ is a feasible dual solution for the metric induced on $S$.
    Therefore, we have
    \begin{equation*}
    K(S) \geq w(T_0\hull{S}) \geq w(T\hull{S}) - 2D. \qedhere
    \end{equation*}
\end{proof}

Let $T_0$ denote the MST shortening satisfying $w(T_0) \geq K - D$.
This MST shortening plays a key role in our analysis.
Eventually, after minor modifications, it yields another MST shortening $T^*$, which we refer to as the \emph{comparison tree}.
The edges of $T^*$ are divided into two parts:
(1)~a~core subtree, representing the \enquote{long} parts of the metric and
(2)~a~hanging forest, consisting of low-height subtrees hanging from the core, representing the rest of the metric.
We shall compare the local solutions in $\Fdual$ with the core and hanging parts of $T^*$, to show that $w(\Fdual)$ is close to $M$ when $D \leq \epsilon M$.
We emphasize that $T_0$ and $T^*$ are never constructed by the algorithm, and they are only defined for the sake of the analysis.
\section{The Core}

This section covers the construction of the core.
On a high level, the algorithm starts with a carefully chosen set of terminals $\Sbase \subseteq V$.
These terminals break down the MST into fragments, which the algorithm is able to explore locally.
From there, the algorithm repeatedly refines $\Sbase$ by repeatedly adding $r$-nets to it, for progressively smaller radii, yielding a final terminal set $S$.

We compute the optimal dominated Steiner tree $C$ on the metric induced by $S$, and refer to it as the \emph{computed core}.
Each vertex $v \in V$ is then mapped to a point $q_v \in \RC$, the continuous extension of $C$, which allows us to decompose the metric into cells in the next section.

Going through this section, we modify the MST shortening $T_0$ step by step to obtain a final shortening $T^*$, which is used to analyze $w(\Fdual)$.
We prove a \emph{core alignment lemma}, stating that the metric induced on $S$ by $T^*$ and the one induced by $C$ are the same up to small perturbations.

\subsection{Base Terminals and the MST Fragments}


We obtain two vertex sets $\Sroot$ and $\Ssplit$, where $\Sbase = \Sroot \cup \Ssplit$, and use them to define the MST fragments as follows:
Let $\Froot$ be the minimum spanning forest rooted at $\Sroot$.\footnote{The algorithm does not compute $\Froot$, but conceptually it could be obtained using the Kruskal algorithm, except if the two endpoints of an edge are already connected to roots in $\Sroot$, the edge is ignored.}
This already divides the MST into subtrees after discarding some edges.
The subtrees are then further split into fragments as follows.
Let $\Fsplit := \Froot\hull{\Sroot\cup\Ssplit}$ include $\Sroot$ and edges lying on the paths from vertices in $\Ssplit$ to their root in $\Sroot$.
The fragments are obtained from the connected components of $\Froot$ after removing the vertices of $\Fsplit$.
For each component, we also include its unique attachment edge to $\Fsplit$, which serves as the root of the fragment.
Its endpoint in $\Fsplit$ is not included in the fragment’s vertex set
We use $\mc{F}$ to denote the collection of the fragments.

Here, we only state the properties of $\Sbase$.
In \cref{sec:implementation}, we explain how $\Sbase$ is actually computed.

\begin{restatable}[Fragment decomposition]{lemma}{fragmentdecomposition}
    \label{lem:fragment-decomposition}
    For a sufficiently small constant $a_0$, we can obtain $\Sroot$ and $\Ssplit$ using $\Ot(n^{2 - a_0})$ queries such that, with high probability, the following properties hold:
    \begin{enumerate}
        \item 
        $
            \card{\Sbase} \leq \card{\Sroot} + \card{\Ssplit} = \Ot(n^{1-a_0}).
        $

        \item Every fragment $F\in\mc{F}$ contains at most $\Ot(n^{a_0})$ vertices.

        \item Given any vertex $v$, the algorithm can recover the unique fragment that includes $v$, or determine that no such fragment exists, using $\Ot(n^{1+5a_0})$ queries.
        This includes recovering all its vertices and edges, and the distances from every vertex of the fragment to every vertex in $V$.
    \end{enumerate}
\end{restatable}

\subsection{Core Terminals}

We compute a terminal set $S \subseteq V$ whose hull serves as a skeleton capturing the long parts of the metric.
For a tree $T$ and a nonempty set $S \subseteq V(T)$, let $T\hull{S}$ denote the minimal subtree of $T$ spanning $S$.

Starting from $S=\Sbase$ and $a=a_0$, we repeatedly enlarge $S$, maintaining $\card{S}=\Ot(n^{1-a})$.
Whenever a vertex is added to $S$, the algorithm queries its distances to all other vertices.
Let $K(S)$ denote the value of \eqref{eq:dual} on the metric induced by the terminal set $S$.
It can be shown that $K(S) \leq K \leq M$ (we prove this in \Cref{cor:ks-leq-k}). 

If, at any point, $K(S)>(1+\epsilon)\hM/2$, we already have the required certificate that $K$ is large. In this case, the algorithm may return with $\hL=K(S)$ and stop.
Otherwise, the algorithm sets
\begin{equation*}
    r=\hM n^{-1+a/10},
\end{equation*}
and computes an $r$-net $N \subseteq V$ in the metric.
We remark $r$-net has a size $\card{N} = O(\hM / r) = O(n^{1 - a/10})$,
since there exists a Hamiltonian cycle of length at most $O(\hM)$.
The algorithm can compute the $r$-net greedily using $O(\card{N} \cdot n) = O(n^{2 - a/10})$ queries:
While there exists a vertex $v$ with $d(v, N) > r$,
add it to $N$ and query all its distances.
Then, the algorithm checks if adding $N$ to $S$ sufficiently increases $K(S)$. That is, if
\begin{equation*}
    K(S\cup N)-K(S) \geq \epsilon\hM,
\end{equation*}
then $S$ is replaced by $S\cup N$, and $a$ is replaced by $a/10$, and the algorithms repeats.

This stopping condition allows us to have a core subtree $T_0\hull{S}$, that can reach every part of the metric with little cost.
More precisely, let the \emph{net expansion} $E_N = T_0\hull{S \cup N} \setminus T_0\hull{S}$ be the set of edges between the subtree of $T_0$ spanning $S$ and the one spanning $S \cup N$.
We further shorten $T_0$ by setting the weight of the edges in $E_N$ to $0$, yielding a new tree $T_1$.
As a result, the non-zero parts of $T_1$ can be divided into two groups:
\begin{enumerate}
    \item the core part $T_1\hull{S}$ and
    \item subtrees hanging from $T_1\hull{S \cup N}$, which have a height of at most $r$ since $N$ is an $r$-net and $T_0$ is dominated by the metric.
\end{enumerate}
The stopping condition ensures that this shortening can be done at only a small cost:

\begin{lemma}
\label{lem:core-terminals}
Assuming that $K(S)$ remains smaller than $(1 + \epsilon)\hM/2$, the algorithm uses $\Ot(n^{2-a/10})$ queries, for some constant $a>0$,
and produces sets $S\supseteq \Sbase$ and $N$ satisfying
\begin{equation*}
    \card{S}=\Ot(n^{1-a}),\qquad
    \card{N}=O(n^{1-a/10}),\qquad
    K(S\cup N)-K(S)\leq \epsilon\hM,
\end{equation*}
where $N$ is an $r$-net for $r=\hM n^{-1+a/10}$.
The new tree $T_1$, obtained by shortening $E_N = T_0\hull{S \cup N} \setminus T_0\hull{S}$, satisfies
\begin{equation*}
    w(T_0)-w(T_1)\leq 2D + \epsilon\hM,
\end{equation*}
and therefore,
\begin{equation*}
    w(T_1)\geq K - 3D - \epsilon\hM.
\end{equation*}
\end{lemma}
\begin{proof}
Each iteration increases $K(S)$ by at least $\epsilon\hM$.
Since $K(S)\leq K\leq M\leq\hM$, there are at most $1/\epsilon$ iterations.
Therefore, starting from $a=a_0$ and replacing $a$ by $a/10$ after each enlargement, the final value of $a$ is also a constant.
Since, by assumption, the early-return condition $(1 + \epsilon)\hM/2$ is never met, the algorithm terminates with
\begin{equation}
    K(S\cup N)-K(S)\leq\epsilon\hM. \label{eq:terminal-stab}
\end{equation}

At termination, we have $\card{S} = \Ot(n^{1-a})$ and $\card{N} = O(\hM / r) = O(n^{1 - a/10})$.
The queries of the algorithm consist of row queries at $S \cup N$.
Hence, their number is at most $\Ot(n^{2 - a/10})$.

It remains to bound $w(T_0) - w(T_1)$.
Recall that $T_1$ was obtained from $T_0$ by setting the weight of the edges in the net expansion $E_N = T_0\hull{S \cup N} \setminus T_0\hull{S}$ to zero.
Therefore,
\begin{align*}
    w(T_0)-w(T_1)
    &=w(T_0\hull{S\cup N})-w(T_0\hull{S})\\
    &\leq w(T\hull{S\cup N})-w(T\hull{S})\\
    &\leq K(S\cup N)+2D-K(S)\\
    &\leq 2D+\epsilon\hM.
\end{align*}
Here, the first inequality follows from the fact that $T_0$ is a shortening of $T$;
the second inequality is due to \cref{lem:k-vs-hull};
and the last inequality holds due to the stopping condition of the algorithm \eqref{eq:terminal-stab}.
Finally, the lower bound on $w(T_1)$ follows from $w(T_0)\geq M - 2D$, which concludes the proof.
\end{proof}

\subsection{The Computed Core}

The subtree $T_1\hull{S}$ provides a skeleton for the comparison tree.
However, it is defined only for purposes of the analysis, and the algorithm does not have access to it.
Instead, the algorithm computes an optimal dominated Steiner tree $C$ for the metric induced by the core terminals $S$.
That is, it holds that
\begin{equation*}
    w(C)=K(S).
\end{equation*}
Since all terminal distances have already been queried, $C$ can be computed without additional queries.
We refer to $C$ as the \emph{computed core}.
It represents the \enquote{long} parts of $T_1$, i.e.\ $T_1\hull{S}$.
More formally, in the next section we argue that the distances induced on $S$ by $T_1$ and the ones induced by $C$ are the same up to small perturbations.

Next, the algorithm assigns each vertex $v\in V$ a location $q_v$ on $\RC$, where $\RC$ denotes the continuous extension of $C$:


\begin{lemma}
\label{lem:q-map}
For any dominated tree $C$ spanning $S$, the distances $d(v,s)$, for $v\in V$ and $s\in S$,
suffice to compute a map $q:V\to \RC$ satisfying
\begin{equation*}
    q_s=s \quad(s\in S)
    \qquad \text{and} \qquad
    d_C(q_u,q_v)\leq d(u,v) \quad(u,v\in V).
\end{equation*}
Although this condition involves all pairs of vertices, knowing the distances to the terminals $S$ suffices to construct the map.
\end{lemma}
\begin{proof}
We begin by describing the mapping $q$.
Consider the possible locations for $q_v$ that respect the distances between $v$ and the terminals:
\begin{equation*}
    I_v=\{x\in \RC : d_C(x,s)\leq d(v,s)\text{ for every }s\in S\}.
\end{equation*}
This set can be expressed as the intersection of balls $B^v_s$, where $q_v$ must be in $B^v_s$ so that $d(v, s)$ dominates $d_C(q_v, s)$:
$$
I_v = \bigcap_{s \in S}B^v_s, 
\qquad \text{where}\ 
B^v_s = \{x \in \RC : d_C(x, s) \leq d(v, s)\}.
$$
Note that both $I_v$ and $B^v_s$ may include interior points on the edges of $C$.

We argue that $I_v$ is non-empty.
Take any two terminals $s, t \in S$.
Since $C$ is dominated by the metric, we have that
$$
d_C(s, t) \leq d(s, t) \leq d(v, s) + d(v, t).
$$
That is, the length of the $C$-path from $s$ to $t$ is at most the sum of the radii of $B^v_s$ and $B^v_t$.
Therefore, $B^v_s \cap B^v_t \neq \emptyset$ for all $s, t \in C$.
By \cref{lem:tree-helly}, the intersection of all the balls, which is equal to $I_v$, is also nonempty.

For any $v$, we let $q_v$ be the center of $I_v$.\footnote{It can be seen that the center is unique, but that is not necessary for our argument.}
Here, the center $q_v$ is the point in $I_v$ that minimizes,
$$
\max_{u \in I_v} d_C(q_v, u),
$$
and this value is referred to as the radius of $I_v$.
Observe that $I_v$ is defined by the distances to the terminals $S$.
Furthermore, for a terminal $s$, the set $I_s$ consists of the single point $s$, and hence $q_s = s$.

It remains to show that $q$ satisfies the non-expansive property.
Take two vertices $u, v \in V$, and let $h = d(u, v)$.
First, we show that every point in $I_u$ is within distance $h$ of $I_v$.
Take $x \in I_u$ and a terminal $s \in S$. Then, it holds that
$$
d_C(x, s) \leq d(u, s) \leq h + d(v, s),
$$
where the first inequality follows from $x \in I_u \subseteq B_s^u$ and the second is the triangle inequality.
Put differently, the ball of radius $h$ around $x$ intersects with every ball $B^v_s$. Therefore, by \cref{lem:tree-helly}, it also intersects with $I_v$.

Now, let $\rho_u$ and $\rho_v$ be the radii of $I_u$ and $I_v$ respectively, and assume $\rho_u \geq \rho_v$.
Take $x$ and $y$ to be the endpoints of a diameter in $I_u$. Their midpoint is $q_u$.
One of these endpoints, say $x$, is in the direction away from $q_v$, meaning:
$$
d_C(x, q_v) = d_C(x, q_u) + d_C(q_u, q_v) = \rho_u + d_C(q_u, q_v).
$$
Take the point $y \in I_v$ that is closest to $x$. Then, by the triangle inequality, it holds that:
$$
d_C(x, y) \geq d_C(x, q_v) - d_C(y, q_v) \geq \rho_u + d_C(q_u, q_v) - \rho_v \geq d_C(q_u, q_v),
$$
which concludes the proof of non-expansiveness.
\end{proof}

\begin{corollary} \label{cor:ks-leq-k}
    Given a vertex set $S \subseteq V$, it holds that 
    $$
    K(S) \leq K.
    $$
\end{corollary}
\begin{proof}
    Take an optimal dominated Steiner tree $C$ spanning $S$, so that $w(C) = K(S)$, and let the mapping $q$ be as in \cref{lem:q-map}.
    We can construct a Steiner tree for the entire metric by placing each vertex $v$ at $q_v$.
    Formally, the Steiner tree is constructed by breaking down the edges of $C$ at the mapping points (i.e., at the image of $q$),
    and attaching each vertex $v$ to its mapping point $q_v$ with an edge of weight $0$.
    The resulting tree has the same weight as $C$.
    It remains to show that it is dominated by the (entire) metric, which follows from the non-expansion property of $q$:
    \begin{equation*}
    d_C(q_u, q_v) \leq d(u, v). \qedhere
    \end{equation*}
\end{proof}

\subsection{Core Alignment and the Comparison Tree}

We now relate the computed core $C$ to the core subtree $T_1\hull{S}$.
That is, we show that after small perturbations, $C$ and $T_1\hull{S}$ agree on the distances for any two vertices $u, v \in T_1\hull{S}$.
These perturbations come in two forms:
\begin{enumerate}
    \item A set of segments $B \subseteq \RC$ which are set to zero. These segments may include only part of an edge.
    We use $C^* = C/B$ to denote the resulting tree and refer to it as the \emph{aligned core}.
    \item A further shortening of $T_1$, which yields the final comparison tree $T^*$.
\end{enumerate}
We emphasize that $C^*, T^*, T_0$, and $T_1$ are only defined for the purposes of the analysis, when proving $\Fdual$ provides a certificate of a large $K$.
The algorithm ever computes only $C$ and the mapping $q$.

\begin{lemma}[Core Alignment]
\label{lem:core-alignment}
There exists a collection of core segments $B$ and a dominated shortening $T^*$ of $T_1$ such that:
\begin{enumerate}
    \item $w(T^*\hull{S}) = w(C^*) = K(S)-w(B)$. \label{item:alignment-1}
    \item The distances in $C^*$ are dominated by $T^*$. That is, for all $u,v\in V$, it holds that
    \begin{equation*}
        d_{C^*}(q_u,q_v)\leq d_{T^*}(u,v).
    \end{equation*}
    Furthermore, this holds with equality for vertices in the core subtree $T_1\hull{S}$.
    \label{item:alignment-2}
    \item Every edge outside $T_1\hull{S}$ has the same weight in $T^*$ as it does in $T_1$.
    \label{item:alignment-3}
\end{enumerate}
The total length set to zero in $C$ and the additional shortening of $T_1$ satisfy
\begin{equation*}
    w(B)\leq 6D+\epsilon\hM
    \qquad \text{and} \qquad
    w(T_1)-w(T^*)\leq 8D+\epsilon\hM.
\end{equation*}
The latter additionally implies that
\begin{equation*}
    w(T^*)\geq K-11D-2\epsilon\hM.
\end{equation*}
\end{lemma}
\begin{proof}
We construct $B$ in two steps.
First, for each $e=(u,v)\in T\hull{S}$, consider the path from $q_u$ to $q_v$ in $C$, denoted by $P_e$.
Since $T\hull{S}$ spans $S$ and the leaves of $C$ are members of $S$, the union of these paths covers $C$.
Since $q$ is non-expansive, it holds that
\begin{equation}
    w(P_e)=d_C(q_u,q_v)\leq d(u,v)=w_T(e).\label{eq:path-dom}
\end{equation}
Hence, the path weights sum up to at most $w(T\hull{S})$.

Let $B_0$ consist of all segments covered by more than one path.
Then, it holds that
\begin{equation*}
    w(B_0)
    \leq \left(\sum_{e\in T\hull{S}}w(P_e)\right)-w(C)
    \leq w(T\hull{S})-K(S)
    \leq 2D,
\end{equation*}
where the last inequality follows from \cref{lem:k-vs-hull}.
At this point, let $B=B_0$ and set the weights of these segments to zero in $C$.

Second, process every edge $e=(u,v)\in T$.
If the current path weight $w_C(P_e)$ exceeds $w_{T_1}(e)$, add segments of $P_e$ to $B$ and set their weights to zero until it equals $w_{T_1}(e)$.
Recall that before this modification the path is at most $w_{T}(e)$ (see \eqref{eq:path-dom} above).
Therefore, as a result of it, the weights decrease by at most
\begin{equation*}
    w_T(e)-w_{T_1}(e).
\end{equation*}

Let $C^*=C/B$.
As a result of the second step, for every $e = (u, v) \in T$, it holds that
\begin{equation}
    d_{C^*}(q_u,q_v)\leq w_{T_1}(e).
    \label{eq:alignment-edge-bound}
\end{equation}
Moreover, by \cref{lem:core-terminals}, it holds that
\begin{align*}
    w(B)
    &\leq w(B_0)+\sum_{e\in T}
        \bigl(w_T(e)-w_{T_1}(e)\bigr)\\
    &\leq 2D+4D+\epsilon\hM
    =6D+\epsilon\hM.
\end{align*}

Now, we define $T^*$, a shortening of $T_1$:
\begin{equation*}
    w_{T^*}(uv)=
    \begin{cases}
        d_{C^*}(q_u,q_v), & uv\in T\hull{S},\\
        w_{T_1}(uv), & uv\notin T\hull{S}.
    \end{cases}
\end{equation*}
This is a shortening of $T_1$ by \eqref{eq:alignment-edge-bound}, and hence it is also dominated by the metric.
It trivially satisfies property~\ref{item:alignment-3} in the lemma statement.

To see property~\ref{item:alignment-1} holds,
observe that every positive-length segment of $C^*$ is covered by exactly one of the paths $P_e$ for $e\in T\hull{S}$.
Consequently, summing the weights $w_{T^*}(e)$ counts every surviving core segment exactly once.
Hence,
\begin{equation*}
    w(T^*\hull{S})=w(C^*)=K(S)-w(B).
\end{equation*}

To see property~\ref{item:alignment-2} holds, first consider vertices $u,v\in T\hull{S}$.
Consider appending the paths $P_e$ for $e$ on the $u$-$v$ path in $T^*$.
This gives a walk on $C^*$ that starts from $q_u$ and ends at $q_v$.
Any edge in this walk that is off the $q_u$-$q_v$ path in $C^*$ must be in $B_0$, and hence has weight $0$.
Therefore, it holds that
\begin{equation*}
    d_{T^*}(u,v) = \sum_{e\in P_{T^*}(u, v)}w_{C^*}(P_e)=d_{C^*}(q_u,q_v).
\end{equation*}
For arbitrary $u,v\in V$, the definition of $T^*$ and \eqref{eq:alignment-edge-bound} give
\begin{equation*}
    d_{C^*}(q_u,q_v)
    \leq \sum_{(x,y)\in P_{T^*}(u, v)} d_{C^*}(q_x,q_y)
    \leq \sum_{(x,y)\in P_{T^*}(u, v)} w_{T^*}(x, y)
    =d_{T^*}(u,v).
\end{equation*}

Finally, to analyze $w(T^*)$, observe that only the edges in $T_1\hull{S}$ are shortened. Hence,
\begin{align*}
    w(T_1)-w(T^*)
    &=w(T_1\hull{S})- w(T^*\hull{S})\\
    &\leq w(T\hull{S})-(K(S)-w(B))\\
    &\leq 8D+\epsilon\hM,
\end{align*}
where the last inequality follows from \cref{lem:k-vs-hull} and the bound on $w(B)$.
Combining this with $w(T_1)\geq K-3D-\epsilon\hM$ yields
\begin{equation*}
    w(T^*)\geq K-11D-2\epsilon\hM. \qedhere
\end{equation*}
\end{proof}

\section{Cell Decomposition and Local Solutions}
We partition the core $C$ into cells.
Then, we compute a local solution (i.e., a dominated Steiner tree) for each, and finally merge them to get a feasible dual solution $\Fdual$. 
We emphasize that $\Fdual$ is not explicitly computed by the algorithm; rather, its size is estimated using sampling and local exploration.

Call a vertex of $C$ \emph{essential} if its degree is not two. This includes the leaves.
Choose a uniformly random radius $R \in [\ell, 2\ell]$, where
$$
\ell = \hM/n^{1 - a/2}.
$$
Take the balls of radius $R$ around the essential vertices, and consider their union.
These balls are subsets of the continuous extension $\RC$ and may include parts of the edges.
The connected components of this union form the \emph{buffer cells}.
The parts of $\RC$ not belonging to a buffer cell
form edge-disjoint paths.
Each such path is divided into randomly shifted \emph{interval cells} of length $\ell$.
More precisely, for each such path, take a uniformly random shift parameter $\theta \in [0, \ell]$.
Then, the endpoints of the intervals are of the form $\theta + j\ell$, for integers $j$, except the first and the last.

For a cell $P \subseteq \RC$, let $V_P$ be the set of vertices that are mapped to $P$, that is, vertices $v$ such that $q_v \in P$.
Let $\partial P$ denote its \emph{ports}, which are the points shared with other cells.
An interval cell has two ports, while a buffer cell may have one or more.

For each cell $P$, we require a local solution $t$, which is a Steiner tree spanning $V_P \cup \partial P$, that satisfies the \emph{cell distance constraints}:
\begin{align}
    d_{t}(u,v)&\leq d(u,v)
        &&(u,v\in V_P),\nonumber\\
    d_{t}(v,p)&\leq d_C(q_v,p)
        &&(v\in V_P,\ p\in\partial P),\nonumber\\
    d_{t}(p,p')&\leq d_C(p,p')
        &&(p,p'\in\partial P).
    \label{eq:cell-port-constraints}
\end{align}
The first condition is the standard domination constraint imposed by the metric.
Additionally, we require the $t$-distances to and between the ports to be dominated by the $C$-distances.
This ensures the local solutions can be merged.



\begin{claim} \label{clm:join-locals}
    Given local solutions $t_P$ for each cell $P$ satisfying \eqref{eq:cell-port-constraints},
    we can combine them to obtain a global solution $F$, which is a Steiner tree spanning $V$ and dominated by the metric.
\end{claim}
\begin{proof}
Consider appending all the local solutions together.
That is, attach the local solutions at their shared ports to obtain a Steiner tree $F$ spanning $V$.
To see that $F$ is dominated by the metric, take two vertices $u, v \in V$.
Consider the path from $q_u$ to $q_v$ on $C$.
Suppose it goes through cells $P_0, P_1, \ldots, P_k$, and ports $p_1, p_2, \ldots, p_k$, where the port $p_i$ is shared between $P_{i-1}$ and $P_{i}$ (see \cref{fig:core-cell-decomposition}).
Then, we have:
\begin{align*}
    d_F(u, v)
    &= d_F(u, p_1) + \sum_{i} d_F(p_i, p_{i+1}) + d_F(p_k, v) \\
    &\leq d_C(q_u, p_1) + \sum_{i} d_C(p_i, p_{i+1}) + d_C(p_k, q_v) \\
    &= d_C(q_u, q_v) \\
    &\leq d(u, v).\qedhere
\end{align*}
\end{proof}

\section{Construction of $\Fdual$}
\label{sec:construction}

We construct a local tree $t_P$ for each cell $P$, satisfying the cell distance constraints \eqref{eq:cell-port-constraints}.
Gluing these trees at their shared ports gives the dominated Steiner tree $\Fdual$.
In \cref{sec:analysis}, we analyze $L = w(\Fdual)$, showing that $L \approx M$ when $K \approx M$.
In \cref{sec:implementation}, we show how the algorithm estimates $L$.

We divide the cells into types: contracted and regular.
For an interval cell $P$, let $m_P$ denote the weight of its port-rooted minimum spanning forest.
This is the minimum weight of a forest spanning $V_P\cup\partial P$ with exactly one port in each component.
Edges between vertices $u,v\in V_P$ have weight $d(u,v)$, and edges from a vertex $v$ to a port $p$ have weight $d_C(q_v,p)$.

A cell $P$ is \emph{contracted} if any of the following conditions holds:
\begin{enumerate}
    \item $P$ is a buffer cell.
    \label{item:regular-interval}
    \item $P$ is an interval cell with $\card{V_P}\geq n^a$, referred to as a \emph{highly-populated} cell.
    \label{item:regular-population}
    \item $P$ is an interval cell with $m_P>Q\ell$, where $\ell = \hM/n^{1 - a/2}$ and $Q$ is a sufficiently large constant, referred to as an \emph{MSF-heavy} cell.
    \label{item:regular-forest-length}
\end{enumerate}
All remaining cells are \emph{regular}.

For each regular cell $P$, let the local solution $t_P$ be a maximum-weight Steiner tree spanned by $V_P\cup\partial P$ satisfying \eqref{eq:cell-port-constraints}.
Write $K_P=w(t_P)$.
The algorithm explicitly computes these local solutions.
The rest of this section describes the local solutions in contracted cells.

\subsection{Contracted Cells}

For a contracted cell $P$, we construct $t_P$ by shortening MST fragments and joining them at a common root.
After joining them at the root, we need to further shorten the weights to ensure that their combination is also dominated by the metric (and also satisfies \cref{eq:cell-port-constraints}).
We do so through the minimum fractional vertex cover in an auxiliary \emph{conflict graph}.
This defines the local solution for $P$.
The algorithm estimates its contribution without explicitly constructing it.

\subsubsection{Cell Fragments}

Consider the collection of MST fragments defined by $\Sroot$ and $\Ssplit$.
We further delete some vertices to obtain smaller fragments whose vertices all belong to the same cell, and refer to them as \emph{cell fragments}.

Formally, let $\mc{F}$ denote the collection of the MST fragments, and let $P$ be any contracted cell.
We say a vertex $v \in V_P$ is \emph{guarded} if $d_C(q_v, p) \leq \lambda r$ for some port $p \in \partial P$, where $\lambda$ is a sufficiently large constant.
Let $V_P^\circ$ denote the unguarded vertices:
\begin{equation*}
    V_P^\circ
    =
    \{v\in V_P:
      d_C(q_v,p)>\lambda r
      \text{ for every }p\in\partial P\}.
\end{equation*}
For cell $P$, we obtain the cell fragments $\mc{F}_P$ by considering the MST fragment subgraphs induced by $V_P^\circ$.
That is, for $F \in \mc{F}$, any connected component of $F[V_P^\circ]$ forms a cell fragment of $P$ (this includes singleton components).

The root $o_F$ of a cell fragment $F\in\mc{F}_P$ is its closest vertex to its original root in the MST fragment.

\subsubsection{Local Shortenings and Discretization}
Let $P$ be a contracted cell.
For each cell fragment $F\in\mc{F}_P$, take a shortening $t_F$ of maximum weight that satisfies
\begin{align}
    d_{t_F}(u,v)&\leq d(u,v)
        &&(u,v\in V(F)),\nonumber\\
    d_{t_F}(v,o_F)&\leq d(v,x)
        &&(v\in V(F),\ x\in V\setminus V(F)),\nonumber\\
    d_{t_F}(v,o_F)&\leq\lambda r
        &&(v\in V(F)).
    \label{eq:contracted-constraints}
\end{align}
Here, the shortening is defined with respect to the MST weights.
That is, each edge $e \in F$ is assigned a weight between zero and $w_T(e)$.
The first condition is the standard domination constraint imposed by the metric, the second bounds distances to vertices outside the fragment, and the third bounds its depth.

We further discretize $t_F$ by replacing each edge of weight $w$ with $\floor{w/\sigma}$ edges of weight $\sigma$,
where $\sigma = \epsilon \hM / 100 n$ is the discretization scale.
Note that since the weights are being rounded down, the discretized trees still satisfy \eqref{eq:contracted-constraints}.
Finally, define
\begin{equation*}
    A_P=\sum_{F\in\mc{F}_P}w(t_F).
\end{equation*}

\subsubsection{The Conflict Graph}

The conflict graph $\Gamma_P$ has one vertex for each segment of the discretized trees.
Two edge segments $e,f$ belonging to different cell fragments are adjacent in $\Gamma_P$ if there exist vertices $u, v$ in their respective fragments such that $e$ is the $i$-th segment on the path from $u$ to its fragment root, 
$f$ is the $j$-th segment on the path from $v$ to its fragment root, and
\begin{equation}
    (i+j)\sigma>d(u,v).
    \label{eq:segment-conflict}
\end{equation}
There are no conflicts between segments of the same fragment.

Let $\tau_P$ be the weight of the minimum fractional vertex cover in the conflict graph:
\begin{equation*}
    \tau_P
    =
    \sigma\min\left\{
        \sum_e y_e:
        0\leq y_e\leq1,\quad
        y_e+y_f\geq1
        \text{ for every conflict }(e,f)
    \right\}.
\end{equation*}
Take an optimal half-integral cover $y$, with $y_e\in\{0,\frac{1}{2},1\}$.
To obtain the local solution $t_P$, 
we join all the discretized shortenings at their roots to form a common root $o_P$,
and shorten the segments according to the vertex cover.
That is, the weight of each edge segment $e$ is decreased by $\sigma y_e$.
Additionally, for every vertex in
\begin{equation*}
    V_P\setminus\bigcup_{F\in\mc{F}_P}V(F),
\end{equation*}
we add a zero-weight edge joining it to $o_P$ (this includes the guarded vertices and vertices of $\Fsplit$ that belong to $V_P$).
Finally, we include the ports in the tree, and connect every port to $o_P$ by an edge of weight zero.

\begin{claim}
\label{clm:contracted-cell-feasibility}
The tree $t_P$ satisfies \eqref{eq:cell-port-constraints} and has weight $A_P-\tau_P$.
\end{claim}
\begin{proof}
First, we prove that we can assume that, in every cell fragment, the cover values are non-increasing going down the fragment and away from the root.
Take two edge segments $e$ and $f$ such that $e$ is an ancestor of $f$.
Observe that every neighbor of $f$ is also a neighbor of $e$.
Therefore, if $y_f > y_e$, we can swap their values, and the resulting cover would still be feasible.
Repeating this gives a cover of the same total weight with non-increasing values down the fragments.

Consequently, along a path starting at a fragment root,
the cover values occur in the order $1,\frac{1}{2},0$.
For the path starting at the root and ending at vertex $v$, let $a_v$ and $b_v$ be the total lengths of the segments with values $\frac{1}{2}$ and $0$ (before applying the cover).
Its distance to the root after applying the cover is
\begin{equation*}
    d_{t_P}(v,o_P)=b_v+\frac{a_v}{2}.
\end{equation*}

Consider two vertices $u$ and $v$ in different fragments.
Take the highest segment $e$ above $u$ that has a value $y_e = 0$,
and the highest segment $f$ above $v$ that has a value
$y_f \leq \frac{1}{2}$.
Assuming $e$ and $f$ exist, since the sum of their values is less than $1$ and $y$ is a cover, there is no conflict between $e$  and $f$.
By \eqref{eq:segment-conflict},
we obtain
\begin{equation*}
    b_u+b_v+a_v\leq d(u,v).
\end{equation*}
If either does not exist, the same inequality follows from the outside-vertex (i.e., second) constraint in \eqref{eq:contracted-constraints}.
Similarly, we can show that:
\begin{equation*}
    b_u+a_u+b_v\leq d(u,v),
\end{equation*}
and averaging the two inequalities yields
\begin{equation*}
    d_{t_P}(u,v)
    = d_{t_P}(u, o_P) + d_{t_P}(v, o_P)
    = b_u+\frac{a_u}{2}+b_v+\frac{a_v}{2}
    \leq d(u,v).
\end{equation*}
Therefore, after applying the cover, the distances between fragments are dominated by the metric.
Distances within a fragment only decrease, so they remain dominated by the metric, and thus $t_P$ is dominated by the metric of $V_P$.

We also need to show that the port constraints from \eqref{eq:cell-port-constraints} hold.
All ports have distance zero from $o_P$.
Therefore, the constraint between ports is satisfied trivially.
Every guarded vertex has distance zero to $o_P$ and hence to all ports.
Every unguarded vertex $v$ has depth at most $\lambda r$, by \cref{eq:contracted-constraints}.
Hence, we have that
\begin{equation*}
    d_{t_P}(v,p)\leq\lambda r<d_C(q_v,p).
\end{equation*}

Finally, the shortening reduces the total segment weight by exactly $\sigma\sum_e y_e=\tau_P$.
Therefore, $w(t_P)=A_P-\tau_P$.
\end{proof}

\subsection{The Global Solution}

By \cref{clm:join-locals}, we can join the local solutions $t_P$ at their shared ports to obtain $\Fdual$, a global Steiner tree dominated by the metric on $V$.
Its weight is
\begin{equation}
    L=w(\Fdual)
    =
    \sum_{P:\text{ regular}}K_P
    +
    \sum_{P:\text{ contracted}}(A_P-\tau_P).
    \label{eq:dual-solution-value}
\end{equation}
Since $\Fdual$ is a feasible dual solution, we have $L\leq K$.
In the next section, we analyze $L$ and show that $L \approx M$ when $K \approx M$.
\section{Analysis of $w(\Fdual)$}
\label{sec:analysis}

\begin{lemma}
\label{lem:dual-weight}
The weight $L=w(\Fdual)$ satisfies
\begin{align*}
    \E[L]
    \geq &
    w(T^*)-8\bigl(M-w(T^*)\bigr)-12Qw(B)\\
    &-\frac{2M}{Q}-\frac{M}{\lambda}-9n\sigma-o(M).
\end{align*}
Consequently, for sufficiently large constants $Q$ and $\lambda$, sufficiently small constant $\epsilon>0$,
\begin{equation*}
    \E[L]
    \geq
    K-c_1D-c_2\epsilon\hM,
\end{equation*}
where $c_1,c_2>0$ depend only on $\epsilon$.
The expectation is over the random cell decomposition.
The same lower bound also holds with $L - \tau$ on the left-hand side.
\end{lemma}

Fix the core $C$, the map $q$, the alignment segments $B$, and the tree $T^*$ before drawing the cell decomposition.
We compare the local solutions with the core and the subtrees of $T^*$ hanging from $T^*\hull{S\cup N}$.
Observe that each hanging subtree has height at most $r$, due to the fact that $N$ is an $r$-net and $T^*$ is dominated by the metric.

We first break down the hanging subtrees according to the cells.
For every hanging edge $(u,v)$ (meaning an edge outside $T^*\hull{S \cup N}$) whose endpoints belong to different cells, with $v$ the child of $u$, set the weight of every edge on the path from $v$ to the subtree root to zero.
Every remaining positive-weight component then belongs to a single cell.
For a cell $P$, we call these the \emph{hanging parts} of that cell and denote their total weight by $R_P$.

In each following section, we combine and modify the hanging parts and the core to obtain feasible local solutions for each cell (satisfying \eqref{eq:cell-port-constraints} for regular cells and \eqref{eq:contracted-constraints} on each fragment for contracted cells).
For contracted cells, we also construct fractional vertex covers of the conflict graphs to upper bound the cost of combining the fragment shortenings.

\begin{claim}
\label{clm:crossing-loss}
The expected weight lost in this step is $O(Mr/\ell)=o(M)$.
In particular,
\begin{equation*}
    \E\left[\sum_P R_P\right]
    \geq
    w(T^*)-(K(S)-w(B))-o(M).
\end{equation*}
\end{claim}
\begin{proof}
The random radius and shifts give port density $O(1/\ell)$ on the core.
That is, any fixed subtree of $C$ of length $W$, chosen independently of the cell decomposition, contains $O(W/\ell)$ ports in expectation.
Thus, an edge $(u,v)\in T$ crosses cells with probability at most
\begin{equation*}
    O\left(\frac{d_C(q_u,q_v)}{\ell}\right)
    \leq O\left(\frac{w_T(u,v)}{\ell}\right).
\end{equation*}
Each crossing hanging edge causes at most $r$ additional weight to be set to zero, since each hanging subtree has height at most $r$.
Summing over the edges of $T$ bounds the expected loss by $O(Mr/\ell)$.

The edges of $T^*\hull{S\cup N}\setminus T^*\hull{S}$ have weight zero.
By \cref{lem:core-alignment}, the core part $T^*\hull{S}$ has weight $K(S)-w(B)$.
The claim follows.
\end{proof}

\subsection{Regular Cells}

\begin{lemma}
\label{lem:regular-weight}
The total weight of local solutions among regular cells satisfies
\begin{equation*}
    \E\left[\sum_{P:\text{ regular}}K_P\right]
    \geq
    \E\left[
        \sum_{P:\text{ regular}}
        \bigl(w(P)-w(B\cap P)+R_P\bigr)
    \right]
    -12Qw(B)-o(M).
\end{equation*}
\end{lemma}

For a regular cell $P$ and a port $p \in \partial P$, define its \emph{guard segment} as the part of $P$ within distance $6r$ of $p$ in $C$:
\begin{equation*}
    I_p=\{x\in P:d_C(x,p)\leq 6r\}.
\end{equation*}
A vertex $v\in V_P$ is \emph{guarded} if $q_v$ lies in either guard segment.
A regular cell is \emph{damaged} if, for a port $p$,
\begin{equation*}
    w(B\cap I_p)>3r.
\end{equation*}
Otherwise, it is \emph{undamaged}.

For every guarded vertex, set the weight of its path to the hanging subtree root to zero.
Let $G_P$ denote the total weight removed from the hanging parts of $P$ in this step.
The surviving hanging parts have total weight $R_P-G_P$.

\begin{claim}
\label{clm:regular-guard-loss}
The expected loss from guarded vertices satisfies
\begin{equation*}
    \E\left[\sum_{P:\text{ regular}}G_P\right]
    =O(nr^2/\ell)=o(M).
\end{equation*}
\end{claim}
\begin{proof}
A vertex $v$ belongs to a regular cell and is guarded with probability $O(r/\ell)$.
To see this, observe that if $q_v$ is within $C$-distance $6r\leq\ell$ of an essential vertex, then it cannot belong to a regular cell.
Otherwise, the $6r$-neighborhood of $q_v$ is a segment of length at most $12r$, which contains $12r\cdot O(1/\ell)$ ports in expectation.

Each guarded vertex causes at most $r$ weight to be set to zero along its path to the hanging subtree root.
Summing over the $n$ vertices gives expected loss $O(nr^2/\ell)$.
By the choices $r=\hM n^{-1+a/10}$ and $\ell=\hM n^{-1+a/2}$, and since $\hM=\Theta(M)$,
\begin{equation*}
    O(nr^2/\ell)=O(\hM n^{-3a/10})=o(M). \qedhere
\end{equation*}
\end{proof}

\subsubsection{Damaged Cells}

\begin{lemma}
\label{lem:damaged-weight}
Every damaged regular cell satisfies $K_P\geq w(P)$.
Moreover,
\begin{equation*}
    \E\left[\sum_{P:\text{ damaged}}K_P\right]
    \geq
    \E\left[
        \sum_{P:\text{ damaged}}
        \bigl(w(P)+R_P-G_P\bigr)
    \right]
    -12Qw(B).
\end{equation*}
\end{lemma}

To construct a feasible local solution for a damaged cell $P$, take the original core interval and place each vertex $v\in V_P$ at $q_v$.
This tree satisfies \eqref{eq:cell-port-constraints} and has weight $w(P)$, and hence we have $K_P\geq w(P)$.
It remains to bound the surviving hanging weight of damaged cells.
Note that ideally $K_P$ would be as large as $w(P) + R_P$.
\cref{lem:damaged-weight} essentially states that, ignoring the effects of the modifications for guarded vertices, ignoring the hanging weight costs us at most $12Qw(B)$.

\begin{claim}
\label{clm:regular-hanging-msf}
For every regular cell $P$,
\begin{equation*}
    R_P-G_P\leq m_P\leq Q\ell.
\end{equation*}
\end{claim}
\begin{proof}
We build a tree of weight $R_P - G_P$, and argue that any port-rooted spanning forest must have a larger weight.
Consider the surviving hanging parts of $P$.
That is, start with $T^*$ after the crossing and guard modifications,
and consider only the hanging subtrees associated with $P$.
We combine these into a tree by contracting all of the following into a single vertex:
(1) the root of each hanging subtree,
(2) the guarded vertices, and
(3) the ports.

Observe that the resulting tree has weight $R_P - G_P$.
Furthermore, the tree distances are dominated by all the edge costs used to define the port-rooted MSF:
Guarded vertices have depth zero.
Every other vertex has depth at most $r$, while its distance in $C$ to either port exceeds $6r$.

Now, consider the edges in the port-rooted MSF.
For each edge, consider the corresponding path in the tree we constructed.
Since the tree has all the ports contracted into one vertex,
these paths cover the tree.
Therefore, the port-rooted MSF has weight at least $R_P - G_P$,
and $m_P \leq Q\ell$ is by definition of regular cells.
\end{proof}

\begin{claim}
\label{clm:damaged-count}
The expected number of damaged regular cells of length at least $6r$ is at most $12w(B)/\ell$.
Consequently,
\begin{equation*}
    \E\left[
        \sum_{P:\text{ damaged}}(R_P-G_P)
    \right]
    \leq 12Qw(B).
\end{equation*}
\end{claim}
\begin{proof}
Fix an orientation of a maximal degree-two path $I$ of $C$, that is, a direction from one endpoint to the other.
Consider the set of \emph{bad points} for ports.
That is, the points $x$ such that, in the event of $x$ becoming a port,
the guarded segment in the direction we fixed has an overlap of at least $3r$ with $B$.
Since the guarded segments have length $6r$, the set of bad points has total length at most $2w(B \cap I)$.

Summing over both orientations and all maximal paths gives total length at most $4w(B)$.
On any of these paths, the two possible buffer boundaries and the shifted grid give port density at most $3/\ell$.
Hence, the expected number of damaged cells of length at least $6r$ is at most $12w(B)/\ell$.

Each such cell has surviving hanging weight at most $Q\ell$ by \cref{clm:regular-hanging-msf}.
In a cell shorter than $6r$, every vertex is guarded, so $R_P-G_P=0$.
\end{proof}

Combining $K_P\geq w(P)$ with \cref{clm:damaged-count} proves \cref{lem:damaged-weight}.

\subsubsection{Undamaged Cells}

\begin{lemma}
\label{lem:undamaged-weight}
For every undamaged regular cell $P$,
\begin{equation*}
    K_P
    \geq
    w(P)-w(B\cap P)-12r+R_P-G_P.
\end{equation*}
\end{lemma}

To lower bound $K_P$, we construct a feasible local tree satisfying \eqref{eq:cell-port-constraints}.
If $w(P)<6r$, every vertex is guarded, so $R_P-G_P=0$ and we can easily argue $K_P \geq w(P)$, similar to the previous section, to establish the bound.
Hence, assume $w(P)\geq6r$, so both guard segments have length $6r$.
We combine the core interval with the surviving hanging parts.

For each vertex $v \in V_P$, we define an \emph{attachment point} to the core subtree, which is the closest point $x \in T\hull{S \cup N}$ to $v$ in the original metric $d_T$.
This also naturally defines an attachment point for each surviving hanging subtree of $P$, which is the attachment point of its root.
We combine the image of $P$ in the aligned core $C^*$ with the surviving hanging subtrees by attaching the surviving subtree to $q_x \in C^*$.

\begin{claim}
\label{clm:undamaged-attachments}
Every surviving hanging part of $P$ attaches inside the image of $P$ in $C^*$.
\end{claim}
\begin{proof}
Take a vertex $v$ in a surviving hanging subtree.
The alignment property and the height bound give
\begin{equation*}
    d_{C^*}(q_v,q_x)
    \leq d_{T^*}(v,x)
    \leq r.
\end{equation*}
Since $v$ is unguarded, the path in $C$ from $q_v$ to either port contains the guard segment of that port.
Undamagedness guarantees that each guard segment retains a weight of at least $3r$ in $C^*$.
Thus, $q_v$ is at least $3r$ from either port of $P$, and hence $q_x$ lies inside.
\end{proof}

After attaching each surviving subtree to the image of $P$ in $C^*$, we need to add all other vertices of $V_P$ to our local solution.
To do so, for each $v \in V_P$  that is not in a surviving subtree and has attachment point $x$, we place it at $q_x$ on $C^*$.
If $q_x$ is outside of the image of $P$, we place $v$ at the closest port of $P$ to $q_x$.
Finally, we set the weights of both guard segments to zero.
This loses at most $12r$ of core weight.

\begin{claim}
\label{clm:undamaged-port-caps}
The resulting tree is a feasible local solution satisfying \eqref{eq:cell-port-constraints}.
\end{claim}
\begin{proof}
First, we argue that the tree is dominated by the metric.
By \cref{lem:core-alignment} and the zero weights on $T^*\hull{S\cup N}\setminus T^*\hull{S}$, any two attachment points $x,y\in T^*\hull{S\cup N}$ satisfy
\begin{equation*}
    d_{C^*}(q_x,q_y)=d_{T^*}(x,y).
\end{equation*}
Thus, joining the hanging subtrees to $C^*$ at the points $q_x$ preserves distances between their vertices: each path between distinct hanging subtrees has the same core length, and paths within a hanging subtree are unchanged.
The crossing and guard modifications only decrease edge weights, and moving the remaining vertices to the nearest point of the cell interval does not increase distances.
Thus, distances between vertices of $V_P$ remain bounded by the metric.

The port-to-port distance starts at $w_C(P)$ and can only decrease. 
Therefore, it remains to check the vertex-to-port constraints.
That is, $d_t(v,p)\leq d_C(q_v,p)$ for every $v\in V_P$ and port $p$, where $t$ denotes the final local tree.
Fix $v$ and $p$, and let $x$ be the attachment point of $v$.
Let $a=q_x$ if $q_x$ lies in the image of $P$, and otherwise let $a$ be its closest port.
The vertex $v$ either lies at $a$ or is connected to $a$ through its surviving hanging part.

First, we bound the core distance from $a$ to $p$ before the guard segments are set to zero.
The alignment property and the height bound give $d_{C^*}(q_x,q_v)\leq r$.
Since $q_v$ lies in the image of $P$, moving $q_x$ to $a$ cannot increase this distance.
Consequently,
\begin{equation*}
    d_{C^*}(a,p)
    \leq d_{C^*}(a,q_v)+d_{C^*}(q_v,p)
    \leq r+d_C(q_v,p).
\end{equation*}

Next, we bound the core distance that remains after the guard segments are set to zero.
Since $P$ is undamaged, the guard segment at $p$ retains at least $3r$ of weight in $C^*$.
If $a$ lies in this segment, setting its weight to zero makes the distance from $a$ to $p$ zero.
Otherwise, the path from $a$ to $p$ contains the entire segment, so its length decreases by at least $3r$.
Thus,
\begin{equation*}
    d_t(a,p)
    \leq \max\{0,d_{C^*}(a,p)-3r\}
    \leq \max\{0,d_C(q_v,p)-2r\}.
\end{equation*}

Finally, we account for the distance from $v$ to $a$.
If $v$ is guarded, its path to $x$ was set to zero, so $v$ lies at $a$.
The preceding bound therefore gives $d_t(v,p)\leq d_C(q_v,p)$.

If $v$ is unguarded, then $d_t(v,a)\leq r$ and $d_C(q_v,p)>6r$.
Hence,
\begin{equation*}
    d_t(v,p)
    \leq r+\max\{0,d_C(q_v,p)-2r\}
    \leq d_C(q_v,p). \qedhere
\end{equation*}
\end{proof}

The retained core weight in this local solution is at least $w(P)-w(B\cap P)-12r$, and the surviving hanging weight is $R_P-G_P$.
These parts are spanned by the vertices of $V_P$ and the ports.
Since this is a feasible local solution, optimality of $K_P$ proves \cref{lem:undamaged-weight}.

There are $O(M/\ell+\card{S})$ interval cells, so the loss of $12r$ per cell is $o(M)$ in total.
Summing \cref{lem:damaged-weight,lem:undamaged-weight} and applying \cref{clm:regular-guard-loss} proves \cref{lem:regular-weight}.

\subsection{Contracted Cells}

\begin{lemma}
\label{lem:contracted-weight}
The contracted-cell solutions satisfy
\begin{align*}
    \E\left[
        \sum_{P:\text{ contracted}}(A_P-\tau_P)
    \right]
    \geq{}&
    \E\left[\sum_{P:\text{ contracted}}R_P\right]
    -8\bigl(M-w(T^*)\bigr)\\
    &-\frac{M}{\lambda}-9n\sigma-o(M).
\end{align*}
In fact, the same lower bound holds with $A_P-2\tau_P$ on the left.
\end{lemma}

We first bound the hanging weight lost at guarded vertices.
We then construct feasible shortenings on the cell fragments satisfying \eqref{eq:contracted-constraints} to lower bound the total local weight.
Finally, we bound the fractional cover cost of combining those optimal shortenings into the local solutions.

For contracted cells, the guard segments are the parts of the cell within distance $\lambda r$ of its ports~(as in \cref{sec:construction}).
A vertex $v$ is guarded if $q_v$ lies in a guard segment.
We modify $T^*$ as follows.
For every guarded vertex $v$, set the weight of its path to its attachment point (in $T^*\hull{S\cup N}$) to zero, as well as the weight of any other hanging edge incident to $v$.
Let $G_P$ denote the hanging weight removed as the result.

\begin{claim}
\label{clm:contracted-guard-loss}
The expected additional loss satisfies
\begin{equation*}
    \E\left[\sum_{P:\text{ contracted}}G_P\right]
    =O(nr^2/\ell)=o(M).
\end{equation*}
\end{claim}
\begin{proof}
A vertex $v$ belongs to a contracted cell and is guarded with probability $O(r/\ell) = o(1)$.
To see this, observe that if $q_v$ is within $C$-distance $\lambda r$ of an essential vertex, then it cannot be guarded: every port is at distance at least $\ell$ from that essential vertex, and $2\lambda r<\ell$.
Otherwise, if there is no essential vertex within distance $\lambda r$, then the $\lambda r$-neighborhood of $q_v$ is a segment of length at most $2\lambda r$, which contains $O(r/\ell)$ ports in expectation.

Each guarded vertex causes a path to its attachment point of weight at most $r$ to be set to zero.
There are $n$ vertices, so the expected loss from these paths is $O(nr^2/\ell)$.
Each hanging edge also has weight at most $r$, and has a guarded endpoint with probability $O(r/\ell)$.
Therefore, their expected total weight is also $O(nr^2/\ell)$, and the total expected loss is $O(nr^2/\ell)=o(M)$.
\end{proof}

We now compare the optimal fragment shortenings with the hanging weight remaining in contracted cells.
To separate this weight from the rest of the tree, we collect the core, the hanging edges already removed, and the hanging parts of regular cells into a subtree $H$.
The hanging weight we want to preserve will then lie outside $H$.

Start with $H=T\hull{S\cup N}$ and add the following paths and edges of the original MST:
\begin{enumerate}
    \item For every hanging edge $(u,v)$ whose endpoints belong to different cells, with $v$ the child of $u$, add the path from $v$ to its attachment point in $T\hull{S\cup N}$.
    \item For every guarded vertex $v$ in a contracted cell, add its path to its attachment point and every hanging edge incident to $v$.
    \item For every vertex $v$ in a regular cell, add its path to its attachment point.
\end{enumerate}
It can be seen that $H$ is connected.
The first two items include the edges set to zero by the crossing and contracted-cell guard modifications.
The third includes all hanging parts of regular cells and their paths to the core; any portion of such a path outside its regular cell was already set to zero by the crossing modifications.
Thus, the weight outside $H$ is exactly the surviving hanging weight of contracted cells:
\begin{equation*}
    w_{T^*}(T\setminus H)
    =
    \sum_{P:\text{ contracted}}(R_P-G_P).
\end{equation*}

We will use two further properties of $H$.
First, every edge outside $H$ belongs to a contracted-cell fragment.
Second, every component outside $H$ is contained in a subtree hanging from $T\hull{S\cup N}$, and therefore has height at most $r$ in $T^*$.

For comparison with the discretized local shortenings, we also discretize the aligned tree $T^*$, using its weights before the crossing and guard modifications.
Let $U$ be obtained by rounding its weight on each original MST edge down to an integer multiple of $\sigma$.
Let $D_U$ denote the total decrease from the original MST weight to $w(U)$, so
\begin{equation*}
    D_U=M-w(U)
    \leq M-w(T^*)+n\sigma.
\end{equation*}

For each contracted cell $P$, a cell fragment $F\in\mc{F}_P$ is a subtree of the MST $T$.
Its shortening $t_F$ is obtained by computing an optimal shortening of $F$ and then discretizing its edge weights, as in \cref{sec:construction}.
For each edge $e\in F$, let $w_{t_F}(e)$ denote the total weight remaining on $e$ in $t_F$.
Let $A_H$ denote the total weight these shortenings assign to MST edges in $H$:
\begin{equation*}
    A_H=
    \sum_{P:\text{ contracted}}
    \sum_{F\in\mc{F}_P}
    \sum_{e\in F\cap H}w_{t_F}(e).
\end{equation*}
We first show that the shortenings $t_F$ retain almost all the MST weight outside $H$, together with most of $A_H$.

\begin{claim}
\label{clm:contracted-local-loss}
The local shortenings satisfy
\begin{equation*}
    \sum_{P:\text{ contracted}}A_P
    \geq
    w_T(T\setminus H)
    +\left(1-\frac{1}{\lambda}\right)A_H
    -2D_U-n\sigma.
\end{equation*}
\end{claim}
\begin{proof}
We construct feasible shortenings that preserve most of the local weight inside $H$ and restore the weights outside $H$ to those in $U$.
For each cell fragment $F\in\mc{F}_P$ of a contracted cell $P$, start with a copy of $t_F$ and modify its edge weights as follows.
Clamp each edge $e\in F$ to weight $\min\{w_{t_F}(e),w_U(e)\}$.

For each cell fragment $F$ and each MST edge $(v,x)$ with $v\in V(F)$ and $x\notin V(F)$, further shorten the path from $v$ to the root of $F$ until its length is at most $w_U(v,x)$.
The outside-vertex constraint originally bounded this depth by $d(v,x)=w_T(v,x)$.
Thus, this correction costs at most $w_T(v,x)-w_U(v,x)$.
Both $v$ and its fragment root lie in $H$, so all corrected paths lie in $H$.
Each MST edge requires at most two such corrections, one for each endpoint.

To make room for the weight restored outside $H$, multiply all local edge weights inside $H$ by $1-1/\lambda$.
This costs at most $A_H/\lambda$.
Every path from a fragment root to a vertex in $H$ lies entirely in $H$, so its length decreases from at most $\lambda r$ to at most $(\lambda-1)r$.
Now restore all edges outside $H$ to their weights in $U$.
Each component outside $H$ has height at most $r$ in $U$, so every vertex still has depth at most $\lambda r$.

We check the remaining constraints in \eqref{eq:contracted-constraints}.
Every edge has weight at most its weight in $U$, so distances within a fragment are bounded by $d_U$, and hence by $d$.
For a vertex $v$ in a fragment and a vertex $x$ outside it, let $(b,c)$ be the first edge leaving the fragment on the path from $v$ to $x$ in $T$.
The endpoint correction gives $b$ depth at most $w_U(b,c)$.
Its path to the fragment root lies in $H$, so scaling and restoring edges outside $H$ cannot increase this depth.
By the triangle inequality, the resulting distance from $v$ to its fragment root is at most
\begin{equation*}
    d_U(v,b)+w_U(b,c)
    \leq d_U(v,x)
    \leq d(v,x).
\end{equation*}
Thus, the constructed shortenings are feasible.

It remains to bound their total weight.
Clamping inside $H$ costs at most $w_T(e)-w_U(e)$ for each fragment edge $e\in H$.
For each MST edge $e$ joining a cell fragment to a vertex outside it, the endpoint corrections cost at most $2(w_T(e)-w_U(e))$.
Outside $H$, replacing the MST weights by those in $U$ loses $w_T(e)-w_U(e)$ per edge.
These three classes of edges are disjoint, so their total charge is at most $2D_U$.
Including the scaling loss, the constructed shortenings have total weight at least
\begin{equation*}
    w_T(T\setminus H)
    +\left(1-\frac{1}{\lambda}\right)A_H
    -2D_U.
\end{equation*}
The local optima before discretization have at least this total weight.
Rounding them loses at most $n\sigma$, which proves the claim.
\end{proof}

\begin{claim}
\label{clm:contracted-cover-cost}
The fractional cover weights satisfy
\begin{equation*}
    \sum_{P:\text{ contracted}}\tau_P
    \leq \frac{A_H}{2}+3D_U.
\end{equation*}
\end{claim}
\begin{proof}
We construct a fractional vertex cover to upper bound the cost $\sum_P\tau_P$ of combining the optimal shortenings into feasible local solutions.
Start again with copies of the discretized shortenings $t_F$ on the cell fragments $F\in\mc{F}_P$.
Perform the clamping and endpoint corrections from \cref{clm:contracted-local-loss}, without the scaling or restoration.
Their total cost is at most $2D_U$.

Move these deletions toward each fragment root until the deleted segments form a rooted subtree.
This preserves the total deletion cost and can only decrease root distances.
Consequently, every surviving segment represented as the $j$-th segment above a vertex $v$ satisfies
\begin{equation*}
    j\sigma\leq d_U(v,x)
\end{equation*}
for every vertex $x$ outside the fragment.

Independently, in each component outside $H$, clamp the original local lengths to their weights in $U$ and move these deletions toward its attachment point in $H$.
This second deletion set costs at most $D_U$.
For any surviving segment outside $H$, represented as the $i$-th segment above $u$, we then have
\begin{equation*}
    i\sigma\leq d_U(u,a),
\end{equation*}
where $a$ is the attachment point of its component in $H$.

Consider a segment in a different fragment, represented as the $j$-th segment above $v$, that survives both deletion sets.
The point $a$ is outside that fragment or is its root, so
$j\sigma\leq d_U(v,a)$.
The original path from $u$ to $v$ passes through $a$.
Therefore,
\begin{equation*}
    (i+j)\sigma
    \leq d_U(u,a)+d_U(a,v)
    =d_U(u,v)
    \leq d(u,v).
\end{equation*}
Thus, after both deletion sets are removed, no conflict is incident to a segment outside $H$.

Assign cover value one to the union of the deletion sets, value one half to every remaining segment in $H$, and zero to all remaining segments.
This is a fractional vertex cover of total weight at most $3D_U+A_H/2$.
\end{proof}

We can now prove \cref{lem:contracted-weight}.
Subtracting twice the cover bound from the local-weight bound in \cref{clm:contracted-local-loss,clm:contracted-cover-cost} gives
\begin{equation*}
    \sum_{P:\text{ contracted}}(A_P-2\tau_P)
    \geq
    w_T(T\setminus H)-8D_U-\frac{A_H}{\lambda}-n\sigma.
\end{equation*}
Here, $A_H\leq M$, $D_U\leq M-w(T^*)+n\sigma$, and $w_T(T\setminus H)\geq\sum_{P:\text{ contracted}}(R_P-G_P)$.
Taking expectations and applying \cref{clm:contracted-guard-loss} proves the lemma, including its stronger bound for $A_P-2\tau_P$.

It remains to prove \cref{lem:dual-weight} by combining the two types of cells.
We first bound the core length omitted by contracted cells.

\begin{claim}
\label{clm:contracted-core-loss}
The total core length of contracted cells satisfies
\begin{equation*}
    \sum_{P:\text{ contracted}}w(P)
    \leq \frac{2M}{Q}+o(M).
\end{equation*}
\end{claim}
\begin{proof}
We start by bounding the core length for the buffer cells.
Consider a maximal path of $C$, whose endpoints are essential vertices and whose interior points are non-essential.
Since the buffer radius is at most $2\ell$, 
such a path contributes at most $4\ell$ to the core length of buffer cells.
Therefore, since there are $O(\card{S})$ such paths, the total core length for buffer cells is at most $O(\card{S}\ell)=o(M)$.

There are at most $n^{1-a}$ highly-populated interval cells, each of length at most $\ell$.
Their total length is at most $n^{1-a}\cdot \ell = o(M)$.

Any remaining contracted cell is an MSF-heavy interval cell, meaning $m_P \geq Q\ell$.
Each has a length of at most $\ell$.
To bound their number, we argue that $\sum_{P:\text{ interval}}m_P\leq2M$.
We use the MST to construct a collection of port-rooted spanning forests for all the interval cells.
For each such cell $P$, keep all MST edges with both endpoints in $V_P$.
Consider the resulting connected components of $V_P$.
For each component, choose one MST edge $(u,v)$ with $u$ in the component and $v\notin V_P$.
The path from $q_u$ to $q_v$ in $C$ passes through a port $p \in \partial P$.
Connect $u$ to this port at cost
\begin{equation*}
    d_C(q_u,p)
    \leq d_C(q_u,q_v)
    \leq d(u,v).
\end{equation*}
Each component is now connected to exactly one port, so this is a feasible port-rooted spanning forest.
Each original MST edge contributes to at most two cells, one for each endpoint, giving $\sum_{P:\text{ interval}}m_P\leq2M$.
As a result, there are at most $2M/(Q\ell)$ MSF-heavy interval cells.
Their total core length is at most $2M/Q$.
\end{proof}

The aligned core has weight $K(S)-w(B)$, and \cref{clm:contracted-core-loss} bounds the part omitted by contracted cells.
Therefore,
\begin{equation*}
    \sum_{P:\text{ regular}}\bigl(w(P)-w(B\cap P)\bigr)
    \geq K(S)-w(B)-\frac{2M}{Q}-o(M).
\end{equation*}
Adding the bounds for regular and contracted cells from \cref{lem:regular-weight,lem:contracted-weight}, and using \cref{clm:crossing-loss} to account for the hanging weight, gives the first bound in \cref{lem:dual-weight}.

Finally, substitute the alignment bounds from \cref{lem:core-alignment} and use $9n\sigma\leq\epsilon\hM$, giving
\begin{equation*}
    \E[L]
    \geq K-c_1D-c_2\epsilon\hM,
\end{equation*}
where $c_1,c_2>0$ depend only on $\epsilon$.
This proves \cref{lem:dual-weight}.

\section{Subquadratic Implementation}
\label{sec:implementation}

This section contains the implementation details and concludes the proof of \cref{thm:main}.
We start by giving an overview of the algorithm.

\begin{algorithm}
    \caption{Overview of the subquadratic algorithm}

    \SetKwInOut{Input}{Input}
    \SetKwInOut{Output}{Output}
    \Input{A metric $(V, d)$}
    \Output{An estimate $\hL$ for $L = w(\Fdual)$}
    
    Compute the set of base terminals $\Sbase = \Sroot \cup \Ssplit$

    Let $S \gets \Sbase$, and refine it by adding progressively finer $r$-nets

    Compute the optimal dominated Steiner tree $C$ spanning $S$

    Decompose the core into cells
    
    For each regular cell $P$, compute $K_P$, which is the maximum weight of a Steiner tree spanning $V_P \cup \partial P$ that satisfies \cref{eq:cell-port-constraints}.

    Compute estimates $\hA$ and $\htau$, where
    $$
    A=\sum_{P:\text{ contracted}}A_P,
    \qquad
    \text{and}
    \qquad
    \tau=\sum_{P:\text{ contracted}}\tau_P.
    $$
    Here, $A_P$ denotes the sum of the cell fragment shortening weights satisfying \eqref{eq:contracted-constraints},
    and $\tau_P$ is the weight of the minimum fractional vertex cover in the conflict graph used to combine them.

    \Return{$\hL = \max\left\{0,\left(\sum_{P:\text{ regular}}K_P\right)+\hA-2\htau-3\epsilon\hM\right\}$}
\end{algorithm}

\begin{lemma}
\label{lem:implementation-estimate}
For sufficiently small $a, \epsilon>0$, the algorithm uses $\Ot(n^{2-a/10})$ distance queries, and returns an estimate $\hL$ satisfying the following with probability $\frac{2}{3}$:
\begin{enumerate}
    \item $0\leq\hL\leq K$, and
    \item if $D \leq \epsilon M$, then $\hL \geq \frac{1 + \epsilon}{2}\hM$.
\end{enumerate}
\end{lemma}

\noindent
We prove \cref{lem:implementation-estimate} in this section, but first we show how it implies \cref{thm:main}.

\begin{proof}[Proof of \cref{thm:main}]
Return
\begin{equation*}
    \widehat{\TSP}
    =
    \begin{cases}
        2\hM,
        & \text{if }\hL\geq(1+\epsilon)\hM/2,\\
        (2-\epsilon)\hM,
        & \text{otherwise}.
    \end{cases}
\end{equation*}
On the success event of \cref{lem:implementation-estimate}, the first case satisfies $\TSP\geq2K\geq2\hL\geq(1+\epsilon)\hM$, so returning $2\hM$ gives a ratio below two.
In the second case, the lemma implies $D>\epsilon M$, and hence $\TSP\leq M+K=2M-D\leq(2-\epsilon)M$.
Both cases give
\begin{equation*}
    \TSP\leq\widehat{\TSP}\leq(2-\epsilon/2)\TSP.
\end{equation*}
\cref{subsec:subquadratic-time} covers the subquadratic-time implementation.
\end{proof}



\subsection{Base Terminals and Access to MST Fragments}

This section is devoted to the proof of \cref{lem:fragment-decomposition}. We restate it here.

\fragmentdecomposition*

\paragraph{Constructing $\Sroot$.}
We begin by uniformly sampling a set of $n^{1-a_0}$ vertices $S_0$ and query their rows.
This sample set allows us to identify a set of \emph{candidate edges} for the MST.
For a vertex $v \notin S_0$, order its edges by weight increasingly.
The candidate edges of $v$ consist of the prefix up to the first edge $(v, u)$ whose other endpoint $u$ is sampled in $S_0$.
Let $Q_0$ be the union of the candidate edges for all vertices.
Observe that given an edge $(u, v)$, we can assert whether it is a candidate edge by comparing $d(u, v)$ to $d(u, S_0)$ and $d(v, S_0)$.
We remark that with high probability every vertex contributes $\Ot(n^{a_0})$ candidate edges to $Q_0$, and hence $\card{Q_0} = \Ot(n^{1 + a_0})$.
However, the max degree of a vertex in $Q_0$ may be larger than $\Ot(n^{a_0})$.

Next, we filter out a vertex set $H_1$, which includes the high-degree vertices in $Q_0$.
We sample an additional vertex sample $S_1$ of size $\Ot(n^{1-2a_0})$ and query their rows.
For each vertex in $S_1$, place the other endpoints of its candidate edges in $H_1$.
Then, with high probability, it holds that $\card{H_1} = \Ot(n^{1-a_0})$, and $H_1$ contains all vertices whose degree in $Q_0$ exceeds $n^{2a_0}$.
We let $\Sroot$ be the union $S_0 \cup H_1$, and also make row queries at every vertex of $H_1$.

\paragraph{Constructing $\Ssplit$.}
We sample $\ceil{n^{1-a_0}}$ vertices uniformly to form $\Ssplit$, and query their rows.
This completes the construction of $\Sbase=\Sroot\cup\Ssplit$.

To bound the fragment sizes, condition on $\Froot$, and for each vertex $v$ consider the subtree rooted at $v$ in $\Froot$.
If this subtree contains $\Omegat(n^{a_0})$ vertices,
then with high probability it includes a vertex sampled in $\Ssplit$, and hence the subtree of $v$ will be fragmented.
More precisely, recall that a fragment will be rooted at an edge $(u, v)$ with $u \in \Fsplit$ and $v \notin \Fsplit$.
The argument above proves that for such a $v$, the subtree size is $\Ot(n^{a_0})$.

\paragraph{Recovering a Fragment.}
Here, we describe how, given a vertex $v$, we can recover the fragment containing $v$ using $\Ot(n^{1+5a_0})$ queries.
We do so by presenting a sequence of subroutines.
These query complexities do not include the $\Ot(n^{2 - a_0})$ queries, due to the row queries at $\Sroot$.

\begin{claim} \label{clm:implement-path-to-root}
    Given a vertex $v \notin \Sroot$, let $x$ be the root of the subtree in $\Froot$ containing $v$.
    The path from $v$ to $x$ can be computed using $\Ot(n^{1 + a_0})$ queries.
\end{claim}
\begin{proof}
    To do so, we run Prim's algorithm starting with $\{v\}$ and continue until a vertex of $\Sroot$ is reached.
    To implement Prim, we start making a row query at $v$.
    In each step, the smallest outgoing edge of the component can be detected through the row queries we already made.
    We add the edge to the component and make a row query at the new endpoint.
    As such, the query complexity is equal to $n$ times the number of iterations until a vertex of $\Sroot$ is reached.
     
    For correctness, recall that $\Froot$ is the minimum spanning forest rooted at $\Sroot$.
    Therefore, it can be computed by contracting the vertices of $\Sroot$ to one vertex, and then using any MST algorithm, such as Prim's.
    Up to the point where a vertex of $\Sroot$ is reached, this is equivalent to the process described above.

    For the query complexity, consider any vertex $v$.
    Take the order in which the vertices are discovered when Prim's algorithm starts at $\{v\}$.
    Note that this ordering is independent of $\Sroot$.
    Therefore, since $\Sroot$ contains $S_0$, which is a sample of size $n^{1-a_0}$,
    there will be a vertex of $\Sroot$ in the first $\Ot(n^{a_0})$ vertices of the ordering, with high probability.
    Therefore, Prim's algorithm reaches $\Sroot$ within $\Ot(n^{a_0})$ iterations, and the query complexity is $\Ot(n^{1 + a_0})$.
\end{proof}

\begin{claim} \label{clm:implement-Froot-children}
    Given a vertex $v \notin \Sroot$, we can obtain the list of its children in $\Froot$ using $\Ot(n^{1 + 3a_0})$ queries, with high probability.
\end{claim}
\begin{proof}
    We start by making a row query at $v$ and obtaining all the candidate edges adjacent to $v$.
    Recall that an edge $(u, v)$ is a candidate if $d(u, v) \leq d(u, S_0)$ or $d(u, v) \leq d(v, S_0)$.
    Furthermore, since the high-degree vertices have already been filtered out (that is, by adding $H_1$ to $\Sroot$),
    therefore the number of candidate edges adjacent to $v$ is at most $n^{2a_0}$.

    The candidate edges include the edges of $\Froot$.
    To see this, consider computing $\Froot$, which is the minimum spanning forest rooted at $\Sroot$, using Kruskal's algorithm.
    When the algorithm inspects the last candidate edge of a vertex $v$, either the edge is added to the MSF or $v$ is already connected to some root in $\Sroot$.
    Either way, $v$ is connected to a root by the time its last candidate edge is inspected.
    Therefore, when the algorithm inspects an edge $(u, v)$ that is not a candidate edge for $u$ or $v$ it will be ignored,
    since the last candidate edge has been inspected $u$ and $v$, and they are already connected to some root.

    To compute the list of the children of $v$,
    we go over all the $O(n^{2a_0})$ candidate edges $(v, u)$, and check if $u$ is the child of $v$ by recovering its path to the root (as in \cref{clm:implement-path-to-root}).
    This takes a total of $O(n^{2a_0}) \cdot \Ot( n^{1 +a_0}) = \Ot(n^{1 + 3a_0})$.
    It finds all the children of $v$ since the edges of $\Froot$ are contained in the candidate edges, as argued above.
\end{proof}

\begin{claim} \label{clm:fragment-recovery}
    Given a vertex $v$, using $\Ot(n^{1 + 5a_0})$ queries, we can find the fragment containing $v$, or assert that it is not included in a fragment, i.e., $v \in V(\Fsplit)$, with high probability.
\end{claim}
\begin{proof}
    First, we show how to assert whether $v \in \Fsplit$.
    Observe that $v\in\Fsplit$ if and only if $v \in \Sroot$ or a vertex in the subtree of $v$ is in $\Ssplit$.
    Therefore, we can simply explore the subtree of $v$ using a DFS algorithm, where we compute the list of the children for a vertex using $\Ot(n^{1 + 3a_0})$ queries (as in \cref{clm:implement-Froot-children}).
    Condition on $\Froot$.
    Since $\Ssplit$ is a sample of size $n^{1 - a_0}$, with high probability, within $\Ot(n^{a_0})$ steps,
    the DFS algorithm either finds a vertex of $\Ssplit$ (asserting $v \in \Fsplit$),
    or fully traverses the subtree (asserting $v \notin \Fsplit$).
    This takes $\Ot(n^{a_0}) \cdot \Ot(n^{1 + 3a_0}) = \Ot(n^{1 + 4a_0})$.

    If $v \notin \Fsplit$, then we consider the path from $v$ to the root of its subtree, and let $x$ be the lowest vertex in the path whose parent is in $\Fsplit$.
    The fragment of $v$ consists of the subtree of $x$, and the edge from $x$ to its parent.
    The path from $v$ to its root and $x$ can be found using $\Ot(n^{1 +5a_0})$, since the length of the path is at most $\Ot(n^{a_0})$ and the membership test for $\Fsplit$ takes $\Ot(n^{1 + 4a_0})$.
    Then, the subtree of $x$ can be computed using a DFS algorithm similar to the previous paragraph, using $\Ot(n^{1 + 4a_0})$ queries.
\end{proof}

We remark that given a contracted cell $P$ and the entire fragment of a vertex $v$,
its cell fragment in $\mc{F}_P$ can easily be obtained by removing the guarded vertices (whose list is obtainable through $C$ and $q$) and the vertices outside the cell.

\subsection{Estimating the Contributions of Contracted Cells}

Recall that the local solutions in the contracted cells are never explicitly constructed.
We estimate the contributions of all contracted cells to $L=w(\Fdual)$ together through sampling.
That is we obtain estimates $\hA$ and $\htau$ such that, with high probability,
\begin{equation}
    \card{\hA-A}\leq\epsilon\hM,
    \qquad
    \frac{\tau}{2}-\epsilon\hM\leq\htau\leq\tau.
    \label{eq:sampling-guarantees}
\end{equation}
Here, the cell decomposition and the local solutions are all fixed, and the randomness considered for $\hA$ and $\htau$ is the randomness associated with sampling.
Recall that
\begin{equation*}
    A=\sum_{P:\text{ contracted}}A_P,
    \qquad
    \tau=\sum_{P:\text{ contracted}}\tau_P,
\end{equation*}
where $A_P$ is the total weight of the discretized shortenings in $P$, and $\tau_P$ is the weight lost when applying the minimum fractional vertex cover for combining them.


\subsubsection{Estimating $A$}

\begin{claim}
    We can compute an estimate $\hA$, satisfying
    $|\hA-A|\leq\epsilon\hM$
    with high probability, using $\Ot(n^{1 + 7a_0}/\epsilon^2)$ queries.
\end{claim}

\begin{proof}
To estimate $A$, we distribute the contribution of a fragment to $A$ among its vertices and sample these contributions.
For a fragment $F\in\mc{F}_P$, let $a_F$ be the weight of its discretized shortening, and let $w_T(F)$ be its original MST weight.
For a vertex $v$ of $F$, let $e_v$ be its parent edge.
The contribution is distributed proportionally to the weight of $e_v$.
\begin{equation*}
    f_v=\frac{w_T(e_v)}{w_T(F)}a_F.
\end{equation*}
We set $f_v = 0$ if $v$ is not in a fragment.
Observe that the contributions within $F$ sum to $a_F$, and hence $\sum_v f_v = A$.

We use \emph{importance sampling} (see \cref{lem:importance-sampling}) to estimate $\sum_v f_v$,
using $b_v = d(v, \Sroot)$ as an upper bound for $f_v$.
We can establish that $b_v$ is an upper bound as follows:
$$
f_v \leq w_T(e_v) \leq d(v, \Sroot).
$$
Here, the first inequality follows from the fact that $a_F$ is the weight of a shortening, hence $a_F \leq w_T(F)$,
and the second inequality can be proven by considering Kruskal's algorithm:
Let $x$ be the closest vertex to $v$ in $\Sroot$.
By the time that the algorithm finishes inspecting $(x, v)$, it connects $v$ to a (directly or through a path) to a vertex in $\Sroot$.
Therefore, any edge on the path from $v$ to its fragment root has a weight of at most $d(v, \Sroot)$.

Importance sampling samples each vertex $v$ with probability $p_v$, proportional to $b_v$, and uses $f_v/p_v$ as the sample value.
That is, letting $\bb = \sum_v b_v$, then
$$
p_v = \frac{b_v}{\bb}.
$$
A standard application of Hoeffding's inequality implies that for an additive error of $\epsilon \hM$, it suffices to average the sample value for $\Ot(\frac{\bb^2}{\epsilon^2 \hM^2})$ vertices.

This allows us to bound the number of required samples by $\Ot(n^{2a_0}/\epsilon^2)$, using $\bb=\Ot(n^{a_0}M)$ and $M\leq\hM$.
To see $\bb=\Ot(n^{a_0}M)$, note that the metric has a Hamiltonian cycle of length at most $2M$.
Divide the Hamiltonian cycle into subpaths by cutting it at the vertices of $\Sroot$.
For a vertex $v$, we bound $d(v, \Sroot)$, by the length of the subpath that includes it.
Since $\Sroot$ includes a vertex sample of size $n^{1 - a_0}$, with high probability, at least one in every $\Ot(n^{a_0})$ consecutive vertices of the Hamiltonian cycle is in $\Sroot$.
Therefore,
$$
\sum_v b_v = \sum_v d(v, \Sroot) \leq \Ot(n^{a_0}M).
$$

Finally, we remark that given a sampled vertex $v$, using $\Ot(n^{1 + 5a_0})$ queries we can recover its fragment $F$ or assert that it is not in any fragment.
In the first case, we can compute the discretized shortening $t_F$ exactly and obtain $f_v$.
In the second case, we have $f_v = 0$.
Since there are $\Ot(n^{2a_0}/\epsilon^2)$ samples, the total query complexity is $\Ot(n^{1 + 7a_0}/\epsilon^2)$.
\end{proof}

\subsubsection{Estimating $\tau$.}
\begin{claim}
We can compute an estimate $\htau$, satisfying $$ \frac{\tau}{2}-\epsilon\hM\leq\htau\leq\tau $$
with high probability, using $\Ot(n^{5/3 + 7a_0})$ queries.
\end{claim}

The algorithm deals with the cell fragments $\mc{F}_P$ in a cell $P$, where each cell fragment $F \in \mc{F}_P$ comes with a discretized shortening $t_F$.
Recall that a conflict graph is defined for each cell to join these shortenings, with $\tau_P$ denoting the weight of a minimum fractional vertex cover.
To estimate $\tau = \sum \tau_P$, we consider the union of these conflict graphs,
and estimate the maximum matching on the double cover.
Here, the double cover of a graph $G$ is an auxiliary graph that includes two vertices $v_L$ and $v_R$ for vertex $v \in V(G)$, and includes two edges $(u_L, v_R)$ and $(u_R, v_L)$ for each edge $(u, v) \in E(G)$.
For simplicity, we assume that the conflict graph is bipartite, and express the algorithm w.r.t.\ the conflict graph rather than its double cover.

The algorithm proceeds in two phases: (1) a peeling phase, which decreases the maximum degree in the conflict graph, as well as the maximum number of potential conflicts for a segment,
and (2) implementing a sublinear algorithm of \citet*{Beh21} on the sparsified conflict graph, which estimates the size of a randomized greedy maximal matching and $2$-approximates the maximum matching size (also with an additive error specified later).
The main challenge in the implementation is accessing the conflict graph implicitly and efficiently.
We start by setting up this implicit access.

Consider the discretized shortening $t_F$ for a cell fragment $F \in \mc{F}_P$.
Take any edge segment in $t_F$.
We use $(u, i)$ to represent this segment
if it is the $i$-th segment lying on the path from $u$ to its fragment root.
Here, $u$ is any vertex in $F$, and $i$ is a positive integer at most $$\imax = \lambda r /\sigma = O(n^{a/10}),$$ since the shortening $t_F$ has a depth limit of $\lambda r$.
Recall that two segments within the same cell but in different fragments have a conflict if they have representations $(u, i)$ and $(v, j)$ such that 
\begin{equation}
(i + j)\sigma > d(u, v). \label{eq:conflict}
\end{equation}

Observe that an edge segment can have many representations.
The algorithm uses the smallest representation of a segment for the purposes of sampling,
but needs to go through all representations of a segment to find its conflicts.
We define a set of \emph{potential representations}
$$
\U = V \times [\imax],
$$
where a potential representation $(v, i)$ may not correspond to any edge segment if the height of $v$ in $t_F$ is smaller than $i\sigma$.
Additionally, for a segment represented by $(u, i)$ we define a set of potential conflicts which includes $(v, j)$ satisfying \eqref{eq:conflict}, for any $v$ in the same cell as $u$ but in a different fragment, and any $j \in [\imax]$.

\begin{claim} \label{clm:potential-representation}
    Given a potential representation $(v, i)$, we can determine whether it corresponds to any edge segment and, if so, compute all representations of that segment along with all its potential conflicts using $\Ot(n^{1 + 5a_0})$ queries with high probability.
\end{claim}
\begin{proof}
    If $v$ belongs to a fragment $F$, we can compute it using $\Ot(n^{1 + 5a_0})$ queries (\cref{clm:fragment-recovery}).
    This allows us to compute the discretized shortening $t_F$, all the representations for all the edge segments in the discretization,
    as well as all their potential conflicts (by simply checking \eqref{eq:conflict}).
\end{proof}

\paragraph{Peeling Phase.}
The goal of this phase is to bound the maximum degree in the conflict graph by $\Delta = \Ot(n^{1/3})$ and the maximum number of potential conflicts for a segment by $\Ot(n^{1/3 + a/10})$.
Later, we account for the effect of this sparsification on the size of the maximum matching.

Conceptually, the algorithm repeatedly samples pivot edge segments, and removes the pivot along with its neighbors in the conflict graph, until the desired maximum degree is reached.
To implement this, we make $\Ot(\card{\U}/\Delta)$ attempts at sampling pivots.
At any point, we maintain a set of representations $\X$ of removed edge segments.
This set will not include all the representations of a removed segment, so to determine if an edge segment has been removed, we need to compute all its representations and check if any of them are in $\X$.

In each attempt, we sample a uniformly random representation from $\U$.
We check if (1) it corresponds to an actual edge segment, (2) it is the smallest representation of that segment,
and (3) none of the representations corresponding to the edge segment appear in $\X$.
If all conditions are met, we accept the corresponding segment as a pivot and add its representation, along with its potential conflicts, to $\X$.
Due to \cref{clm:fragment-recovery}, with high probability, we can carry out this process using
$$
\Ot(\card{\U}/\Delta) \cdot \Ot(n^{1 + 5a_0}) = \Ot(n^{5/3 + 6a_0})
$$
queries.
Let $X$ denote the set of edge segments that have a representation appearing in $\X$.
As a result of removing $X$ from the conflict graph, the maximum degree falls below $\Delta$ with high probability.

Now, we decrease the number of potential conflicts.
Consider any edge segment $e$ not removed by $X$,
and let $D(e)$ denote the number of its potential conflicts not in $\X$.
Due to the peeling, it holds that
$$
\sum_{e \notin X} {D(e)} \leq \Delta\card{\U}.
$$
Let $U_* = 2\hM/\sigma$ be an upper bound for the total number of segments.
We discard any segment $e$ with
$$
D(e) > \frac{\Delta\card{\U}}{\epsilon U_*} = O(n^{1/3 + a/10}).
$$
Use $X'$ to denote the set of segments discarded in this step.
Observe that after removing $X'$, the size of the maximum matching decreases by at most $\card{X'} \leq \epsilon U_*$, which is negligible for our purposes.
We never compute $X'$ explicitly; we lazily check if a segment appears in $X'$ when necessary.

After these transformations, we can use the bounds on the maximum degree and maximum number of potential conflicts to obtain efficient access to the adjacency lists in the conflict graph:
\begin{claim} \label{clm:implement-conflict-adjacency-list}
    Consider the conflict graph after removing $X$ and $X'$. Given a segment $e$, we can obtain the list of its neighbors in the conflict graph using $O(n^{4/3 + 6a_0})$  queries.
\end{claim}
\begin{proof}
    We first recover all representations of $e$ and its potential conflicts (as in \cref{clm:potential-representation}).
    After excluding the representations in $\X$, at most $\Delta\card{\U}/(\epsilon U_*)$ potential conflicts remain, since $e\notin X'$.
    For each potential conflict, we apply \cref{clm:potential-representation} again to determine whether it represents an actual segment and, if so, recover its representations and potential conflicts.
    This information suffices to test membership in $X$ and $X'$ without further queries.
    We discard nonexistent or removed segments (i.e., members of $X$ and $X'$), and identify the remaining segments by their smallest representations to eliminate duplicates.
    This gives the adjacency list of $e$.
    
    We use $\Ot(n^{1+5a_0})$ per potential conflict, so the total cost is
    \begin{equation*}
        (1+D(e)) \cdot \Ot\left(
        n^{1+5a_0}
        \right)
        =
        \Ot(n^{4/3+6a_0}).
        \qedhere
    \end{equation*}
\end{proof}

\paragraph{RGMM Phase.}
We use the matching size estimator of \citet*{Beh21} on the sparsified conflict graph.
Their algorithm estimates the expected size of a randomized greedy maximal matching by sampling vertices and determining whether they are matched.
\begin{lemma}[{\cite[Theorem~1.3]{Beh21}}]
\label{lem:behnezhad-matching}
Let $G$ be an unweighted graph on $N$ vertices with average degree $\bar d$, and let $\mu(G)$ denote its maximum matching size.
For any $\epsilon>0$, there is an algorithm that uses
$\Ot((\bar d+1)/\epsilon^2)$ time and adjacency-list queries\footnote{An adjacency-list query $(v, i)$ asks for the $i$-th neighbor in the adjacency list of $v$.} and, with high probability, returns an estimate $\hmu$ satisfying
\begin{equation*}
    \frac{\mu(G)}{2}-\epsilon N
    \leq \hmu
    \leq \mu(G).
\end{equation*}
\end{lemma}

Let $G$ denote the bipartite double cover of the union of the conflict graphs for different cells, and let
\begin{equation*}
    H = G \setminus (X\cup X')
\end{equation*}
be the graph remaining after peeling.
We apply all the preceding constructions to $G$, including a left or right copy indicator in each representation.

We have implicit access to $H$.
To sample edge segments (i.e., vertices of $H$), we view it as a graph on $\U$.
Each surviving segment copy is identified with its smallest representation.
All other representations are regarded as isolated vertices.
By \cref{clm:implement-conflict-adjacency-list}, we can obtain the adjacency list of any edge segment. That is, we obtain the entire adjacency list for any segment the algorithm explores, as opposed to simply answering adjacency list queries.

\begin{claim}
Conditioned on the peeling guarantees, we can compute an estimate $\hmu_H$ satisfying
\begin{equation*}
    \frac{\mu(H)}{2}-\frac{\epsilon U_*}{16}
    \leq \hmu_H
    \leq \mu(H)
\end{equation*}
with high probability, using $\Ot(n^{5/3+7a_0})$ distance queries.
\end{claim}

\begin{proof}
We apply the matching-size estimator with
$\epsilon'=\epsilon U_*/(16\card{\U})$.
It uses
\begin{equation*}
    \Ot\left(
        (\Delta+1)
        \left(\frac{\card{\U}}{\epsilon U_*}\right)^2
    \right)
    = \Ot(n^{1/3 + a_0/5}).
\end{equation*}
adjacency-list queries, where recall that $\Delta = \Ot(n^{1/3})$.
Due to
\cref{clm:implement-conflict-adjacency-list},
we need 
$$\Ot(n^{1/3+a_0/5})\cdot\Ot(n^{4/3+6a_0})
    \leq\Ot(n^{5/3+7a_0})$$
distance queries in total.
\end{proof}

\paragraph{Final Estimate $\htau$.}
Let $I$ denote the set of accepted pivots, and let $N(I)$ denote their neighbors in $G$.
Since the peeling is sequential, $I$ is an independent set.

For one execution of the peeling and RGMM phases, define
\begin{equation*}
    Z=\card{N(I)}+2\hmu_H.
\end{equation*}
We truncate $\hmu_H$ to $[0,U_*/2]$.
This preserves the guarantees of $\hmu_H$,
while bounding $Z$ by $2U_*$.

We estimate $\mathbb{E}[Z]$ by averaging $\Ot(1/\epsilon^2)$ independent executions of the two phases.
For each execution, we estimate $\card{N(I)}$ up to an additive error of $\epsilon U_* / 16$,
by sampling $\Ot(\imax) = \Ot(n^{a/10})$ representations from $\U$
and checking if they are the smallest representation of a segment in $N(I)$.
Averaging the resulting $Z$ for each execution yields an estimate $\hZ$ satisfying
\begin{equation*}
    \left|\hZ - \mathbb{E}[Z]\right|
    \leq\frac{\epsilon U_*}{8}
\end{equation*}
with high probability.
Each $N(I)$ estimate uses $\Ot(n^{a/10})$ representation samples, so its query cost is smaller than that of the RGMM phase.
The total query complexity is therefore at most $\Ot(n^{5/3+7a_0})$.

Recall that $G$ is the bipartite double cover, and hence
\begin{equation*}
    \tau=\frac{\sigma}{2}\mu(G).
\end{equation*}
To estimate it, we use
\begin{equation*}
    \htau=
    \frac{\sigma}{4}
    \max\left(0,\widehat Z-\frac{\epsilon U_*}{4}\right).
\end{equation*}

\begin{claim}
With high probability, the estimate satisfies
\begin{equation*}
    \frac{\tau}{2}-\epsilon\hM
    \leq\htau\leq\tau.
\end{equation*}
\end{claim}

\begin{proof}
We show that
\begin{equation}
    \mu(G)-\frac{5\epsilon U_*}{4}
    \leq\mathbb{E}[Z]
    \leq2\mu(G)+\frac{\epsilon U_*}{8}.
    \label{eq:peeling-estimate-expectation}
\end{equation}
The claim follows from algebraic manipulations, given that $U_* = 2\hM / \sigma$ and $|\widehat Z-\mathbb{E}[Z]|\leq\epsilon U_*/8$.

For the lower bound, we remark that $G \setminus N(I)$ has the same maximum matching size as $G \setminus X = G \setminus (I \cup N(I))$, since removing $N(I)$ makes the vertices in $I$ isolated.
Furthermore, removing any $k$ vertices from a graph decreases its maximum matching size by at most $k$.
Therefore, considering removing $N(I)$ and $X'$, it holds that
$$
\mu(H) \geq \mu(G) - \card{N(I)} - \card{X'}.
$$
The lower bound then follows by plugging in
$$\mu(H) \leq 2\hmu_H + \epsilon U_*/8
\qquad \text{and} \qquad 
\card{X'} \leq \epsilon U_*.$$

For the upper bound, fix a minimum vertex cover $W$ of $G$, so that $\card{W}=\mu(G)$.
Let
\begin{equation*}
    x=\card{N(I)\cap W}
    \qquad \text{and} \qquad
    y=\card{N(I)\setminus W}.
\end{equation*}
Since $\hmu_H \leq \mu(H)$ with high probability, and $\hmu_H$ is truncated to $U_*/2$,
letting the probability of failure be $\epsilon/8$, it follows that
$$
\E[\hmu_H] \leq \E[\mu(H)] + \epsilon U_* / 16.
$$
As a result, we have that
\begin{align}
\E[Z]
&= \E[\card{N(I)} + 2\hmu_H] \notag \\
&\leq  \epsilon U_* / 8 +  \E[\card{N(I)} + 2\mu(H)] \notag \\
&\leq  \epsilon U_* / 8 +  \E[(x + y) + 2(\card{W} - x)]  \label{eq:w-x}\\
&=  \epsilon U_* / 8 + 2\mu(G) + \E[y - x] , \notag
\end{align}
where \eqref{eq:w-x} follows from the fact that $W \setminus (X \cup X')$, which has size at most $\card{W} - x$, is a vertex cover for $H$.

Therefore, it suffices to show that $\E[y] \leq \E[x]$.
At each peeling attempt, every remaining vertex has the same probability of being selected as a pivot.
Therefore, each remaining edge with exactly one endpoint in $W$ contributes equally to the expected increases of $x$ and $y$.
Edges with both endpoints in $W$ contribute only to the expected increase of $x$.
There are no edges with both endpoints outside $W$, since $W$ is a vertex cover.
Summing over the attempts gives $\E[y]\leq \E[x]$.
\end{proof}

\subsection{Final Estimate $\hL$}
\label{subsec:final-estimate}

\begin{proof}[Proof of \cref{lem:implementation-estimate}]
The preceding implementation claims, together with \cref{lem:core-terminals}, give a total of $\Ot(n^{2-a/10})$ queries for sufficiently small $a_0$ and $a\leq a_0$.
Computing the regular-cell solutions requires only $O(n^{1+a})$ additional queries, since each regular cell can be processed using $O(\card{V_P}^2)$ queries, and $\card{V_P} \leq n^a$.
We stop and return zero if a subroutine exceeds its stated query budget, counting this as a failure.

If the algorithm halts during terminal refinement, the lemma is satisfied since
$(1 + \epsilon)\hM/2 \leq K(S)\leq K$.
Otherwise, assuming that \eqref{eq:sampling-guarantees} holds, the definition of the returned estimate gives
\begin{equation*}
    0\leq\hL\leq L\leq K
    \qquad \text{and} \qquad
    \hL\geq L-\tau-4\epsilon\hM.
\end{equation*}
This proves the first guarantee.

Now, suppose $D\leq\epsilon M$.
Since \cref{lem:contracted-weight} also holds with $A_P-2\tau_P$, the proof of \cref{lem:dual-weight} gives the same bound for $L-\tau$:
\begin{equation*}
    \E[L-\tau]\geq K-c_1D-c_2\epsilon\hM.
\end{equation*}
Since $L-\tau\leq M$, we can apply Markov's inequality to $M-L+\tau$.
The expectation bound above and $D\leq\epsilon M$ then give $L-\tau\geq3M/4$ with probability at least $5/6$, when $\epsilon$ is chosen sufficiently small.
The lower bound on $\hL$ therefore gives $\hL\geq(1+\epsilon)\hM/2$.
The implementation and sampling fail with probability at most $1/6$, so both guarantees hold with probability at least $2/3$.
\end{proof}

\subsection{Subquadratic Time}
\label{subsec:subquadratic-time}

So far we have proved the algorithm can be implemented using $\Ot(n^{2-a/10})$ queries.
Here, we argue that it can also be implemented in sub-quadratic time.
Throughout the section, we assume $a_0 \geq a > 0$ are sufficiently small constants.

The terminal sets $\Sbase$ and the $r$-nets $N$ can be obtained in time $\Ot(n^{2 - a_0})$ and $\Ot(n^{2-a/10})$, same as their query complexities.

For computing local solutions, since the subproblems involve $\Ot(n^{a_0})$ vertices, we can use standard polynomial-time LP solvers:
Computing the optimal shortening of a fragment that satisfies \eqref{eq:contracted-constraints} can be expressed as an LP with $\Ot(n^{a_0})$ variables and $\Ot(n^{2a_0})$ constraints.
Computing $K_P$ reduces to solving an instance of the Cut LP.
Using a minimum-cut algorithm as a separation oracle, we can obtain an algorithm whose runtime is polynomial in $\Ot(n^a)$.

The main challenge is computing the core.
Given a terminal set $S$, this requires two subroutines:
(1) approximating $K(S)$, and
(2) constructing a dominated Steiner tree $C$ whose weight is sufficiently close to $K(S)$.

The multiplicative weights algorithm of \citet*{ChekuriQ17} approximates the Cut LP and produces an implicit feasible dual solution $y$ with value at least $(1-\epsilon)K(S)$ in $\Ot(\card{S}^2)$ time.
Since $\card{S}=\Ot(n^{1-a})$, this is $\Ot(n^{2-2a})$ time.
During terminal refinement we also apply it to $S\cup N$, whose size is $\Ot(n^{1-a/10})$, giving $\Ot(n^{2-a/5})$ time for these calls.
The error due to the estimation of $K(S)$ is within our tolerance.

To construct $C$, first recover the \emph{pairwise loads} defined as 
\begin{equation}
\rho(u,v)=\sum_{A:(u,v)\in\delta(A)}y_A.
\label{eq:loads}
\end{equation}
The solver represents its cuts using one or two edges of auxiliary trees, allowing these loads to be calculated in $\Ot(\card{S}^2)$ time.
Fix any vertex $z \in S$ and compute a maximum spanning tree on $S\setminus\{z\}$ with edge weights
\begin{equation}
    g_{uv}=\frac{\rho(z,u)+\rho(z,v)-\rho(u,v)}{2}.
    \label{eq:g}
\end{equation}
We construct a Steiner tree by processing the spanning-tree edges in decreasing order.
Initially, the Steiner tree consists of one vertex for each terminal $u \in S$.
For each terminal, we define a depth $h_u := \rho(z, u)$.
Then, when we process an edge $(u, v)$, we connect their components at depth $g_{uv}$.
More precisely, let $r_1$ and $r_2$ be the roots of the subtrees that include $u$ and $v$ respectively (each representing their connected component at the time).
We create a new vertex $r$ in the Steiner tree,
assign it a depth $h_r = g_{uv}$, and connect it to $r_1$ and $r_2$ with weights:
$$
d_C(r, r_1) = h_{r_1} - h_r 
\qquad \text{and} \qquad
d_C(r, r_2) = h_{r_2} - h_r.
$$
The new vertex $r$ becomes the root of the merged component.
After all edges are processed, connect the root $r$ of the component containing $S\setminus\{z\}$ to $z$ with an edge of weight $h_r$.
This conversion also takes $\Ot(\card{S}^2)=\Ot(n^{2-2a})$ time, and yields the desired properties:

\begin{claim}
\label{clm:fast-dual-tree}
The resulting tree $C$ has nonnegative edge weights and satisfies
\begin{equation*}
    d_C(u,v)\leq\rho(u,v)\leq d(u,v)
    \qquad(u,v\in S),
\end{equation*}
and
\begin{equation*}
    w(C)\geq\sum_A y_A.
\end{equation*}
\end{claim}
\begin{proof}
Replacing cuts by their complements when necessary, assume $z\notin A$ whenever $y_A>0$.
Then, by \eqref{eq:loads} and \eqref{eq:g} it holds that
\begin{equation*}
    h_u=\sum_{A:u\in A}y_A,
    \qquad
    \text{and}
    \qquad
    g_{uv}=\sum_{A:u,v\in A} y_A.
\end{equation*}
Therefore, we have 
\begin{equation}
    0\leq g_{uv}\leq\min\{h_u,h_v\}. \label{eq:g-vs-h}
\end{equation}
At each merge, each of the two component roots is either a terminal or was created in an earlier step.
In the first case, the edge weight is non-negative due to \eqref{eq:g-vs-h}, and in the second case because the edges are processed in decreasing order of $g_{uv}$.

Let $R$ denote the maximum spanning tree.
For $u,v\ne z$, every edge on their path in $R$ has weight at least $g_{uv}$, since $R$ has maximum length.
Their components therefore merge at depth at least $g_{uv}$, giving
\begin{equation*}
    d_C(u,v)\leq h_u+h_v-2g_{uv}=\rho(u,v).
\end{equation*}
Also, $d_C(z,u)=h_u=\rho(z,u)$, which concludes the proof of the first property.

To prove the second property, for an edge $(a, b)$ in $C$, express its weight as $h_b - h_a$, where $a$ is the parent vertex.
Let $R[A]$ denote the edges of $R$ with both endpoints in $A$.
It follows that:
\begin{align*}
    w(C)
    &=\sum_{u\ne z}h_u-\sum_{(u,v)\in R}g_{uv}\\
    &=\sum_A y_A\bigl(\card{A}-\card{R[A]}\bigr)
    \geq\sum_A y_A,
\end{align*}
where the last inequality holds since $\card{R[A]}\leq\card{A}-1$.
\end{proof}

\section*{AI Disclosure}

We provided ChatGPT-6 Astra with the problem context and our preliminary
ideas. Building on this context and follow-up conversations, it played a substantial role in
developing parts of the final algorithm. This manuscript was written by the authors, who
have verified the correctness and originality of all content, and
take full responsibility for the results.

\bibliographystyle{plainnat}

\bibliography{references}

@inproceedings{CzumajS04,
  author       = {Artur Czumaj and Christian Sohler},
  title        = {Estimating the Weight of Metric Minimum Spanning Trees
                  in Sublinear-Time},
  booktitle    = {Proceedings of the 36th Annual {ACM} Symposium on
                  Theory of Computing ({STOC} 2004)},
  pages        = {175--183},
  publisher    = {ACM},
  year         = {2004},
  doi          = {10.1145/1007352.1007386},
  url          = {https://doi.org/10.1145/1007352.1007386}
}

@misc{Onak16,
  author       = {Krzysztof Onak},
  title        = {Sublinear.info Open Problem 71: Metric {TSP} Cost
                  Approximation},
  year         = {2016},
  howpublished = {Open Problems in Sublinear Algorithms},
  note         = {Suggested at the Baltimore 2016 workshop},
  url          = {https://sublinear.info/71}
}

@inproceedings{ChenKK20,
  author       = {Yu Chen and Sampath Kannan and Sanjeev Khanna},
  title        = {Sublinear Algorithms and Lower Bounds for Metric
                  {TSP} Cost Estimation},
  booktitle    = {47th International Colloquium on Automata, Languages,
                  and Programming ({ICALP} 2020)},
  editor       = {Artur Czumaj and Anuj Dawar and Emanuela Merelli},
  series       = {Leibniz International Proceedings in Informatics
                  ({LIPIcs})},
  volume       = {168},
  pages        = {30:1--30:19},
  publisher    = {Schloss Dagstuhl -- Leibniz-Zentrum f{\"u}r Informatik},
  address      = {Dagstuhl, Germany},
  year         = {2020},
  doi          = {10.4230/LIPIcs.ICALP.2020.30},
  url          = {https://doi.org/10.4230/LIPIcs.ICALP.2020.30}
}

@inproceedings{BhattacharyaKS23,
  author       = {Sayan Bhattacharya and Peter Kiss and
                  Thatchaphol Saranurak},
  title        = {Sublinear Algorithms for {$(1.5+\epsilon)$}-Approximate
                  Matching},
  booktitle    = {Proceedings of the 55th Annual {ACM} Symposium on
                  Theory of Computing ({STOC} 2023)},
  pages        = {254--266},
  publisher    = {ACM},
  year         = {2023},
  doi          = {10.1145/3564246.3585252},
  url          = {https://doi.org/10.1145/3564246.3585252}
}

@inproceedings{ChenKT23TSP,
  author       = {Yu Chen and Sanjeev Khanna and Zihan Tan},
  title        = {Sublinear Algorithms and Lower Bounds for Estimating
                  {MST} and {TSP} Cost in General Metrics},
  booktitle    = {50th International Colloquium on Automata, Languages,
                  and Programming ({ICALP} 2023)},
  editor       = {Kousha Etessami and Uriel Feige and Gabriele Puppis},
  series       = {Leibniz International Proceedings in Informatics
                  ({LIPIcs})},
  volume       = {261},
  pages        = {37:1--37:16},
  publisher    = {Schloss Dagstuhl -- Leibniz-Zentrum f{\"u}r Informatik},
  address      = {Dagstuhl, Germany},
  year         = {2023},
  doi          = {10.4230/LIPIcs.ICALP.2023.37},
  url          = {https://doi.org/10.4230/LIPIcs.ICALP.2023.37}
}

@inproceedings{BehnezhadRRS24,
  author       = {Soheil Behnezhad and Mohammad Roghani and
                  Aviad Rubinstein and Amin Saberi},
  title        = {Sublinear Algorithms for {TSP} via Path Covers},
  booktitle    = {51st International Colloquium on Automata, Languages,
                  and Programming ({ICALP} 2024)},
  editor       = {Karl Bringmann and Martin Grohe and Gabriele Puppis and
                  Ola Svensson},
  series       = {Leibniz International Proceedings in Informatics
                  ({LIPIcs})},
  volume       = {297},
  pages        = {19:1--19:16},
  publisher    = {Schloss Dagstuhl -- Leibniz-Zentrum f{\"u}r Informatik},
  address      = {Dagstuhl, Germany},
  year         = {2024},
  doi          = {10.4230/LIPIcs.ICALP.2024.19},
  url          = {https://doi.org/10.4230/LIPIcs.ICALP.2024.19}
}

@inproceedings{Beh21,
  author    = {Soheil Behnezhad},
  title     = {Time-Optimal Sublinear Algorithms for Matching and Vertex Cover},
  booktitle = {2021 IEEE 62nd Annual Symposium on Foundations of Computer
               Science (FOCS)},
  pages     = {873--884},
  publisher = {IEEE},
  year      = {2021},
  doi       = {10.1109/FOCS52979.2021.00089},
  url       = {https://doi.org/10.1109/FOCS52979.2021.00089}
}

@book{BLM13,
  author       = {St{\'{e}}phane Boucheron and
                  G{\'{a}}bor Lugosi and
                  Pascal Massart},
  title        = {Concentration Inequalities - {A} Nonasymptotic Theory of Independence},
  publisher    = {Oxford University Press},
  year         = {2013},
  url          = {https://doi.org/10.1093/acprof:oso/9780199535255.001.0001},
  doi          = {10.1093/ACPROF:OSO/9780199535255.001.0001},
  isbn         = {978-0-19-953525-5},
  bibsource    = {dblp computer science bibliography, https://dblp.org}
}

@article{DantzigFJ54,
  author       = {George B. Dantzig and
                  D. Ray Fulkerson and
                  Selmer M. Johnson},
  title        = {Solution of a Large-Scale Traveling-Salesman Problem},
  journal      = {Oper. Res.},
  volume       = {2},
  number       = {4},
  pages        = {393--410},
  year         = {1954},
  url          = {https://doi.org/10.1287/opre.2.4.393},
  doi          = {10.1287/OPRE.2.4.393},
  bibsource    = {dblp computer science bibliography, https://dblp.org}
}

@article{GoemansB93,
  author       = {Michel X. Goemans and
                  Dimitris Bertsimas},
  title        = {Survivable networks, linear programming relaxations and the parsimonious
                  property},
  journal      = {Math. Program.},
  volume       = {60},
  pages        = {145--166},
  year         = {1993},
  url          = {https://doi.org/10.1007/BF01580607},
  doi          = {10.1007/BF01580607},
  bibsource    = {dblp computer science bibliography, https://dblp.org}
}

@article{EdmondsJ73,
  author       = {Jack Edmonds and
                  Ellis L. Johnson},
  title        = {Matching, Euler tours and the Chinese postman},
  journal      = {Math. Program.},
  volume       = {5},
  number       = {1},
  pages        = {88--124},
  year         = {1973},
  url          = {https://doi.org/10.1007/BF01580113},
  doi          = {10.1007/BF01580113},
  bibsource    = {dblp computer science bibliography, https://dblp.org}
}

@inproceedings{ChenKT23Steiner,
  author       = {Yu Chen and
                  Sanjeev Khanna and
                  Zihan Tan},
  editor       = {Nikhil Bansal and
                  Viswanath Nagarajan},
  title        = {Query Complexity of the Metric Steiner Tree Problem},
  booktitle    = {Proceedings of the 2023 {ACM-SIAM} Symposium on Discrete Algorithms,
                  {SODA} 2023, Florence, Italy, January 22-25, 2023},
  pages        = {4893--4935},
  publisher    = {{SIAM}},
  year         = {2023},
  url          = {https://doi.org/10.1137/1.9781611977554.ch179},
  doi          = {10.1137/1.9781611977554.CH179},
  bibsource    = {dblp computer science bibliography, https://dblp.org}
}

@inproceedings{Indyk99,
  author       = {Piotr Indyk},
  editor       = {Jeffrey Scott Vitter and
                  Lawrence L. Larmore and
                  Frank Thomson Leighton},
  title        = {Sublinear Time Algorithms for Metric Space Problems},
  booktitle    = {Proceedings of the Thirty-First Annual {ACM} Symposium on Theory of
                  Computing, May 1-4, 1999, Atlanta, Georgia, {USA}},
  pages        = {428--434},
  publisher    = {{ACM}},
  year         = {1999},
  url          = {https://doi.org/10.1145/301250.301366},
  doi          = {10.1145/301250.301366},
  bibsource    = {dblp computer science bibliography, https://dblp.org}
}

@inproceedings{KarlinKG21,
  author       = {Anna R. Karlin and
                  Nathan Klein and
                  Shayan Oveis Gharan},
  editor       = {Samir Khuller and
                  Virginia Vassilevska Williams},
  title        = {A (slightly) improved approximation algorithm for metric {TSP}},
  booktitle    = {{STOC} '21: 53rd Annual {ACM} {SIGACT} Symposium on Theory of Computing,
                  Virtual Event, Italy, June 21-25, 2021},
  pages        = {32--45},
  publisher    = {{ACM}},
  year         = {2021},
  url          = {https://doi.org/10.1145/3406325.3451009},
  doi          = {10.1145/3406325.3451009},
  bibsource    = {dblp computer science bibliography, https://dblp.org}
}

@inproceedings{KarlinKG22,
  author       = {Anna R. Karlin and
                  Nathan Klein and
                  Shayan Oveis Gharan},
  title        = {A (Slightly) Improved Bound on the Integrality Gap of the Subtour
                  {LP} for {TSP}},
  booktitle    = {63rd {IEEE} Annual Symposium on Foundations of Computer Science, {FOCS}
                  2022, Denver, CO, USA, October 31 - November 3, 2022},
  pages        = {832--843},
  publisher    = {{IEEE}},
  year         = {2022},
  url          = {https://doi.org/10.1109/FOCS54457.2022.00084},
  doi          = {10.1109/FOCS54457.2022.00084},
  bibsource    = {dblp computer science bibliography, https://dblp.org}
}

@inproceedings{GharanSS11,
  author       = {Shayan Oveis Gharan and
                  Amin Saberi and
                  Mohit Singh},
  editor       = {Rafail Ostrovsky},
  title        = {A Randomized Rounding Approach to the Traveling Salesman Problem},
  booktitle    = {{IEEE} 52nd Annual Symposium on Foundations of Computer Science, {FOCS}
                  2011, Palm Springs, CA, USA, October 22-25, 2011},
  pages        = {550--559},
  publisher    = {{IEEE} Computer Society},
  year         = {2011},
  url          = {https://doi.org/10.1109/FOCS.2011.80},
  doi          = {10.1109/FOCS.2011.80},
  bibsource    = {dblp computer science bibliography, https://dblp.org}
}

@inproceedings{ChekuriQ17,
  author       = {Chandra Chekuri and
                  Kent Quanrud},
  editor       = {Chris Umans},
  title        = {Approximating the Held-Karp Bound for Metric {TSP} in Nearly-Linear
                  Time},
  booktitle    = {58th {IEEE} Annual Symposium on Foundations of Computer Science, {FOCS}
                  2017, Berkeley, CA, USA, October 15-17, 2017},
  pages        = {789--800},
  publisher    = {{IEEE} Computer Society},
  year         = {2017},
  url          = {https://doi.org/10.1109/FOCS.2017.78},
  doi          = {10.1109/FOCS.2017.78},
  bibsource    = {dblp computer science bibliography, https://dblp.org}
}

@book{traub2024approximation,
  title={Approximation algorithms for traveling salesman problems},
  author={Traub, Vera and Vygen, Jens},
  year={2024},
  publisher={Cambridge University Press}
}


\end{document}